\documentclass[aip,reprint,nofootinbib,superscriptaddress]{revtex4-1}

\pdfoutput=1

\usepackage{graphicx}
\usepackage{dcolumn}
\usepackage{bm}
\usepackage{amsmath}
\usepackage{amssymb}
\usepackage{braket}
\usepackage{booktabs}
\usepackage{multirow}
\usepackage{setspace}
\usepackage{afterpage}
\usepackage{enumitem}
\usepackage{xr}
\usepackage{pdfpages}

\makeatletter
\AtBeginDocument{\let\LS@rot\@undefined}
\makeatother

\begin{document}

\title{Magnetic Field Dependent Microwave Losses in Superconducting Niobium Microstrip Resonators}

\author{Sangil~Kwon}
\email{kwon2866@gmail.com}
\affiliation{Institute for Quantum Computing, University of Waterloo, Waterloo, Ontario N2L 3G1, Canada}
\affiliation{Department of Physics and Astronomy, University of Waterloo, Waterloo, Ontario N2L 3G1, Canada}

\author{Anita~Fadavi~Roudsari}
\affiliation{Institute for Quantum Computing, University of Waterloo, Waterloo, Ontario N2L 3G1, Canada}

\author{Olaf~W.~B.~Benningshof}
\affiliation{Institute for Quantum Computing, University of Waterloo, Waterloo, Ontario N2L 3G1, Canada}
\affiliation{Department of Physics and Astronomy, University of Waterloo, Waterloo, Ontario N2L 3G1, Canada}

\author{Yong-Chao~Tang}
\affiliation{Institute for Quantum Computing, University of Waterloo, Waterloo, Ontario N2L 3G1, Canada}
\affiliation{Department of Electrical and Computer Engineering, University of Waterloo, Waterloo, Ontario N2L 3G1, Canada}

\author{Hamid~R.~Mohebbi}
\affiliation{High Q Technologies LP, Waterloo, Ontario N2L 0A7, Canada}

\author{Ivar~A.~J.~Taminiau}
\affiliation{Neutron Optics LP, Waterloo, Ontario N2L 0A7, Canada}

\author{Deler~Langenberg}
\affiliation{High Q Technologies LP, Waterloo, Ontario N2L 0A7, Canada}

\author{Shinyoung~Lee}
\affiliation{Samsung Electronics, 1 Samsungjeonja-ro, Hwaseong-si, Gyeonggi-do, 445-330, Republic of Korea}

\author{George~Nichols}
\affiliation{Institute for Quantum Computing, University of Waterloo, Waterloo, Ontario N2L 3G1, Canada}
\affiliation{Department of Physics and Astronomy, University of Waterloo, Waterloo, Ontario N2L 3G1, Canada}

\author{David~G.~Cory}
\affiliation{Institute for Quantum Computing, University of Waterloo, Waterloo, Ontario N2L 3G1, Canada}
\affiliation{Department of Chemistry, University of Waterloo, Waterloo, Ontario N2L 3G1, Canada}
\affiliation{Perimeter Institute for Theoretical Physics, Waterloo, Ontario N2L 2Y5, Canada}
\affiliation{Canada Institute for Advanced Research, Toronto, Ontario M5G 1Z8, Canada}

\author{Guo-Xing~Miao}
\affiliation{Institute for Quantum Computing, University of Waterloo, Waterloo, Ontario N2L 3G1, Canada}
\affiliation{Department of Electrical and Computer Engineering, University of Waterloo, Waterloo, Ontario N2L 3G1, Canada}

\date{\today}

\begin{abstract}
We describe an experimental protocol to characterize magnetic field dependent microwave losses in superconducting niobium microstrip resonators.
Our approach provides a unified view that covers two well-known magnetic field dependent loss mechanisms: quasiparticle generation and vortex motion.
We find that quasiparticle generation is the dominant loss mechanism for parallel magnetic fields.
For perpendicular fields, the dominant loss mechanism is vortex motion or switches from quasiparticle generation to vortex motion, depending on cooling procedures. 
In particular, we introduce a plot of the quality factor versus the resonance frequency as a general method for identifying the dominant loss mechanism.
We calculate the expected resonance frequency and the quality factor as a function of the magnetic field by modeling the complex resistivity.
Key parameters characterizing microwave loss are estimated from comparisons of the observed and expected resonator properties.
Based on these key parameters, we find a niobium resonator whose thickness is similar to its penetration depth is the best choice for X-band electron spin resonance applications.
Finally, we detect partial release of the Meissner current at the vortex penetration field, suggesting that the interaction between vortices and the Meissner current near the edges is essential to understand the magnetic field dependence of the resonator properties.
\end{abstract}

\maketitle

\section{Introduction}
\label{sec:intro}

Superconducting resonators have been studied for half a century and their importance has grown especially rapidly in the past decade, driven by increased interest in quantum information and quantum devices.\cite{zmuidzinas, lancaster, hein, haroche, schoelkopf, clarke, gu, wallquist, poot, houck, daniilidis, aspelmeyer, morton, xiang}
Recently, there is a renewed interest in using superconducting resonators for magnetic resonance and therefore to use them in magnetic fields.\cite{morton, xiang, kurizki, ghirri, staudt2012, benningshof2013, malissa2013, probst2013, huebl2013, sigillito2014, grezes2014, tkalcec2014, putz2014, wisby2014, ghirri2015, grezes2015, zollitsch2015, bienfait2016a, bienfait2016b, bonizzoni2016, wisby2016, eichler2017, astner2017}
Our interest is to develop a resonator for X-band electron spin resonance (ESR) of thin films.
We desire to have a small mode volume and a homogeneous microwave magnetic field over the sample, and so we employ a microstrip geometry.\cite{mohebbi2014}
Such resonators have the potential to significantly increase the signal-to-noise ratio, if the resonator maintains a high quality factor in a modest DC magnetic field.

Maintaining a high quality factor is not straightforward in a magnetic field because of magnetic field dependent microwave losses.\cite{mohebbi2014, frunzio2005, song2009b, bothner2011, bothner2012a, bothner2012b, deGraaf2012, deGraaf2014, singh2014, samkharadze2016, ebensperger2016, tang2016, tang2017, bothner2017}
The focus of this paper is to develop experimental methods to understand and characterize the magnetic field dependent loss mechanisms, quasiparticle generation and current-induced motion of vortices.

The quasiparticle loss induced by a magnetic field is determined by both the film quality (clean/dirty) and its thickness. 
Regarding the film quality, dirtier films have a higher Ginzburg--Landau (GL) parameter;\cite{tinkham}
therefore, they survive in higher field.
However, dirtier films have more scattering sites that makes them lossier.
As for the film thickness, thinner films are less sensitive to a magnetic field parallel to the film because the Meissner current does not repel all of the penetrating magnetic field.\cite{tinkham}
Another advantage of thin films is that, if the thickness of a thin film is comparable to or thinner than its GL coherence length, vortices are not easily created by a magnetic field parallel to the film.
However, when the film is too thin its quality degrades because the surface oxide layer and lattice mismatch between the substrate and the film become more important.\cite{gubin2005, lemberger2007}

There are two approaches to avoiding loss from current-induced vortex motion:
one is to suppress the vortex motion of existing vortices and the other is to shift vortex penetration to a higher field.
Most studies on vortices in planar resonators have focused on reducing vortex motion, by introducing artificial pinning sites such as slots\cite{song2009b} or antidots.\cite{bothner2011, bothner2012a, bothner2012b}
The other approach is enhancing a surface barrier which delays vortex penetration until the external field reaches a value above the lower critical field of the resonator.
At this field, called the vortex penetration field, the surface barrier is fully suppressed.\cite{matsushita, kuznetsov1999, brandt2013}
In this work, we focus on the role of the Bean--Livingston surface barrier at the edges of microstrips, \cite{bean1964} rather than on artificial pinning sites.

\begin{figure*}
\centering
\includegraphics[scale=0.5]{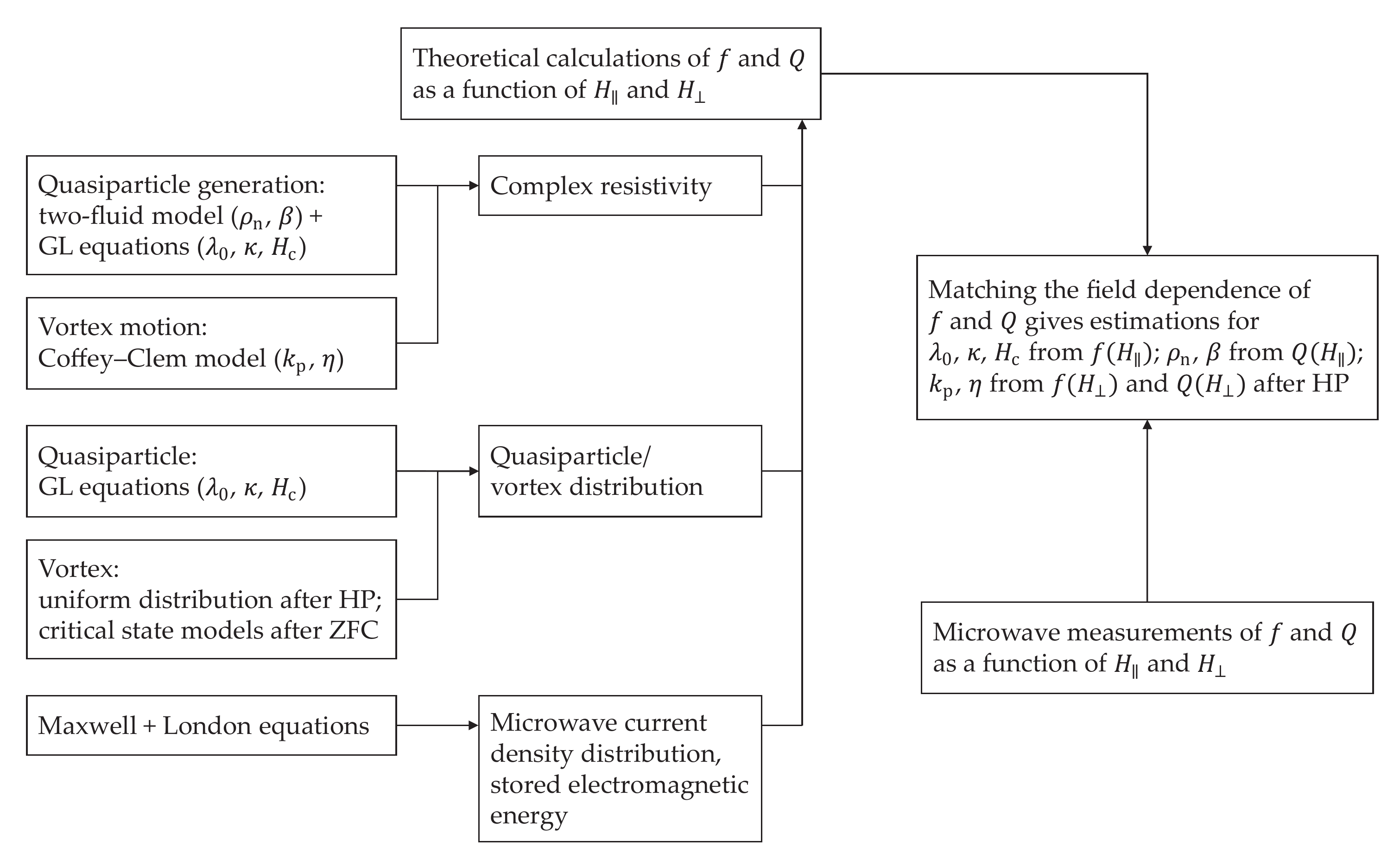}
\caption{\label{fig:flowChart}
Flow chart of our approach to characterize microwave losses induced by DC magnetic fields.
Loss parameters, the parameters for modeling the complex resistivity, are in parentheses in the leftmost boxes.
These are
the zero-field penetration depth of the strip $\lambda_0$,
the GL parameter $\kappa$,
the thermodynamic critical field $H_\textrm{c}$,
the residual resistivity $\rho_\textrm{n}$,
the exponent for the fraction of normal electrons in context of the two-fluid picture $\beta$ (see Sec.~\ref{sec:tf}),
the restoring force constant of a pinning potential $k_\textrm{p}$, and
the viscous drag coefficient associated with vortex motion $\eta$ (see Sec.~\ref{sec:vm}).
ZFC and HP refer to zero-field cooling and heat-pulsing, respectively (see Sec.~\ref{sec:method}).
}
\end{figure*}

Our approach to understanding the loss is outlined in Fig.~\ref{fig:flowChart}.
We systematically characterized a set of superconducting niobium microstrip resonators with different film quality, film thickness, and strip width by measuring their resonance frequencies $f$ and quality factors $Q$ as a function of magnetic fields both parallel to the microwave current $H_\parallel$ and perpendicular to the film $H_\perp$.

We theoretically calculate $f$ and $Q$ for each resonator as a function of magnetic field using standard models for the complex resistivity of superconductors:
the two-fluid model incorporated with the time-dependent GL equations (when quasiparticle generation is the dominant loss mechanism) or the Coffey--Clem model (when vortex motion is dominant).
By varying the parameters, which are used to model the complex resistivity (the parameters in parentheses in the leftmost boxes of Fig.~\ref{fig:flowChart}), we match the theoretical $f$ and $Q$ to determine the loss parameters.

To calculate $f$ and $Q$, the quasiparticle/vortex distribution, microwave current density distribution, and stored electromagnetic energy need to be calculated.
The quasiparticle distribution is given by the time-dependent GL equations used for the complex resistivity;
the vortex distribution is assumed to be uniform or to follow the critical state models, depending on cooling procedures.
The microwave current density distribution and the electromagnetic energy are calculated using Maxwell's equations and the London equations.

An essential step is to identify the dominant loss mechanism, either quasiparticle generation or vortex motion, for each experimental condition.
Once the dominant loss mechanism is known, we can use the appropriate model to compute the complex resistivity.
In this work, we introduce a plot of $Q$ vs. $f^{-2}$, which represents the characteristic relation between the real and imaginary parts of the complex resistivity, as a general method for identifying the dominant loss mechanism (Sec.~\ref{sec:qfPlot}).

One outcome of the approach described in this paper is that we observe an anomaly in the magnetic field dependence of the resonance frequency and interpret it as partial release of the Meissner current along the strip edges at the vortex penetration field, a phenomenon which has not been reported for superconducting resonators (Sec.~\ref{sec:freqAnomaly}).

Finally, our approach allows us to propose design criteria for high quality factor planar resonators that are suitable for ESR applications (Sec.~\ref{sec:conclusion}).

This paper is organized as follows.
Section~\ref{sec:theory} introduces the theories used for the calculations.
Section~\ref{sec:method} describes the details of the resonators and the experimental conditions.
Section~\ref{sec:result} presents results and analysis.
Section~\ref{sec:conclusion} concludes the paper.
Some technical details have been deferred to Supplementary Materials.

\section{Theory}
\label{sec:theory}

\subsection{Resonance Frequency and Quality Factor}
\label{sec:freqQ}

Consider a microstrip line oriented along the $z$ axis with its width along the $x$ axis and thickness along the $y$ axis.
The dissipated power per unit length $P_\textrm{diss}$ as a function of external magnetic field $H$ is
\begin{equation} \label{eq:dissE}
P_\textrm{diss}(H) = \frac{1}{2} \int_\textrm{sc} \rho_1(x,y,H) |J_\textrm{mw}(x,y,\lambda(H))|^2 dxdy
\end{equation}
where ``sc'' stands for ``inside superconducting media'',
$\rho$ is the complex resistivity $\rho_1 + \textrm{i} \rho_2$,
$J_\textrm{mw}$ is the microwave current density, and
$\lambda$ is the magnetic field penetration depth.

There are many other sources of power loss, such as coupling to external circuits or two-level systems.\cite{zmuidzinas, mohebbi2014, goppl2008, sage2011, goetz2016}
We assume that these losses do not have a magnetic field dependence.

The stored electromagnetic energy per unit length $U_\textrm{em}$ can be divided into two parts, the energy stored as an electromagnetic field $U_\textrm{field}$ and the additional energy contribution $U_\textrm{add}$:
\begin{align}
U_\textrm{em}(H)
&= U_\textrm{field}(H) + U_\textrm{add}(H)  \nonumber\\
&= \frac{1}{2} \int_\textrm{all} \mu_0 |H_\textrm{mw}(x,y,\lambda(H))|^2 dxdy \nonumber\\
&\quad + \frac{1}{2} \int_\textrm{sc} \frac{\rho_2(x,y,H)}{\omega} |J_\textrm{mw}(x,y,\lambda(H))|^2 dxdy \label{eq:magE}
\end{align}
where $\mu_0$ is the vacuum permeability, $H_\textrm{mw}$ is the microwave magnetic field strength, and $\omega/2\pi$ is the frequency of an applied electromagnetic field.

The quality factor provides a convenient measure of the loss as $Q^{-1}$:
\begin{equation}\label{eq:Q}
\frac{1}{Q(H)} = \frac{P_\textrm{diss}(H)}{\omega U_\textrm{em}(H)}
\approx \frac{P_\textrm{diss}(H)}{\omega U_\textrm{field}(H)}.
\end{equation}
We make an approximation for the last term as $U_\textrm{field} \gg U_\textrm{add}$.
Therefore, the magnetic field dependence of $\rho_1$ can be studied via measuring $Q$ as a function of $H$.

In a microstrip resonator, the resonance frequency indicates the phase velocity of the microwave signal, which is proportional to $\sqrt{L}$, where $L$ is the effective inductance per unit length.
The quantity $L$ is defined by $U_\textrm{em} = L|I|^2/2$, where $I$ is the total current.
Like $U_\textrm{em}$, $L$ has two terms:
\begin{equation*}
L(H) = L_\textrm{field}(H) + L_\textrm{add}(H),
\end{equation*}
where $L_\textrm{field}$ is the magnetic inductance from $U_\textrm{field}$, and $L_\textrm{add}$ is an additional inductance from $U_\textrm{add}$. Hence,
\begin{align}
L_\textrm{field}(H) &= \frac{1}{|I|^2} \int_\textrm{all} \mu_0 |H_\textrm{mw}(x,y,\lambda(H))|^2 dxdy, \label{eq:magInduc}\\
L_\textrm{add}(H) &= \frac{1}{|I|^2} \int_\textrm{sc} \frac{\rho_2(H)}{\omega} |J_\textrm{mw}(x,y,\lambda(H))|^2 dxdy. \label{eq:addInduc}
\end{align}

Given the assumption that the capacitance of a microstrip resonator is independent of magnetic field, the magnetic dependent part of $L$ can be measured by the following equation:
\begin{equation}\label{eq:freqL}
\frac{f^{-2}(H) - f_0^{-2}}{f_0^{-2}} = \frac{L(H) - L_0}{L_0},
\end{equation}
where $f_0$ is the resonance frequency at zero-field, and $L_0$ is the effective inductance at zero-field.
Hence, the magnetic field dependence of $\rho_2$ can be extracted from $f^{-2}(H)$.

The discussions so far suggest that we need $\rho$, $J_\textrm{mw}$, and $U_\textrm{field}$ to calculate $f^{-2}/f_0^{-2}$ and $Q^{-1}$ as a function of $H$.
Among these, we can simulate $J_\textrm{mw}$ and $U_\textrm{field}$ by solving Maxwell's equations and the London equations (see Sec.~S2).
In the next subsection, we introduce several models for $\rho$.

\subsection{Magnetic Field Dependent Loss Mechanisms}

\subsubsection{Quasiparticle Generation}
\label{sec:tf}

When $\omega$ is low enough that $\omega \tau_\textrm{qp} \ll 1$, where $\tau_\textrm{qp}$ is the quasiparticle scattering time, the two-fluid model provides a convenient description of the complex conductivity due to the quasiparticle generation $\sigma_\textrm{tf,1}-\textrm{i}\sigma_\textrm{tf,2}$,\cite{tinkham}
\begin{align}
\sigma_\textrm{tf,1} &= \frac{n_\textrm{n}}{n_\textrm{tot}} \sigma_\textrm{n}, \label{eq:con1TF} \\ 
\sigma_\textrm{tf,2} &= \frac{n_\textrm{s} {e_\textrm{s}}^2}{m_\textrm{s}\omega} 
= \frac{1}{\omega\mu_0\lambda^2}, \label{eq:con2TF}
\end{align}
where $n_\textrm{n}$ is the local number density of normal electrons (quasiparticle), 
$n_\textrm{s}$ is the local number density of superconducting electrons (Cooper pair), 
$n_\textrm{tot}$ is the total number density of conduction electrons,
$\sigma_\textrm{n}$ is the inverse of $\rho_\textrm{n}$,
$e_\textrm{s}$ is the charge of a superconducting electron,
$m_\textrm{s}$ is the mass of a superconducting electron,
and $\lambda$ is the penetration depth.
The corresponding complex resistivity $\rho_{\textrm{tf}}$ is given by $\rho_{\textrm{tf},i} = \sigma_{\textrm{tf},i} / (\sigma_\textrm{tf,1}^2 + \sigma_\textrm{tf,2}^2)$.
As $\sigma_{\textrm{tf},1} \ll \sigma_\textrm{tf,2}$ for the most of the magnetic field range, $\rho_{\textrm{tf},1}$ is approximately proportional to $\sigma_\textrm{n} \lambda^4$.
Hence, for dirty superconductors, the number of scattering sites affects $\rho_{\textrm{tf},1}$ chiefly via their effect on $\lambda$.\cite{tinkham}

In this work, $n_\textrm{s}$ is calculated using the time-dependent GL equations in terms of the complex order parameter $\psi$:
\begin{equation}\label{eq:nsGL}
n_\textrm{s}(x,y,H) = |\psi(x,y,H)|^2.
\end{equation}
As the GL theory does not give $n_\textrm{n}$, we introduce an empirical expression for $n_\textrm{n}$ with an additional exponent $\beta$:
\begin{equation}\label{eq:nnGL}
\frac{n_\textrm{n}(H)}{n_\textrm{tot}} = \left[1 - \frac{n_\textrm{s}(H)}{n_\textrm{s}(0)}\right]^{\beta}.
\end{equation}
This expression fits our data well (see Sec.~\ref{sec:lossPara}).
For $\beta > 1$ ($\beta < 1$), $n_\textrm{n}$ is less (greater) than that would be predicted by an ideal two-fluid model, $\beta = 1$.
Thus the loss parameters associated with quasiparticle generation are $\sigma_\textrm{n}$(=$1/\rho_\textrm{n}$), $\beta$, and parameters for the GL equations, $\lambda_0$, $\kappa$, and $H_\textrm{c}$ (see Sec.~S1 for details on the GL equations).

\subsubsection{Vortex Motion}
\label{sec:vm}

Current-induced vortex motion is an important source of microwave dissipation.
To describe the complex resistivity associated with vortex motion, we consider the interactions between pinning potentials and vortices.
Among the several accepted models for this,\cite{coffey1991, brandt1991, dulcic1993b, pompeo2008} we use the complex resistivity based on the Coffey--Clem model $\rho_\textrm{CC,1}+\textrm{i}\rho_\textrm{CC,2}$ given by \cite{silva2006}
\begin{equation} \label{eq:CCbasic}
\rho_{\textrm{CC},i} \approx \rho_{\textrm{tf},i} + \rho_{\textrm{vm},i},
\end{equation}
where $\rho_\textrm{vm}$ is the complex resistivity due to vortex motion. 
A useful property of Eq.~\eqref{eq:CCbasic} is that the total complex resistivity is the sum of $\rho_\textrm{tf}$ and $\rho_\textrm{vm}$.
Here, $\rho_\textrm{vm}$ is given by \cite{pompeo2008, silva2006}
\begin{equation}\label{eq:resCCvm}
\begin{split}
\rho_\textrm{vm,1} &= \rho_\textrm{ff}\frac{(\omega/\omega_\textrm{eff})^2+\epsilon}{1+(\omega/\omega_\textrm{eff})^2}, \\
\rho_\textrm{vm,2} &= \rho_\textrm{ff}\frac{\omega/\omega_\textrm{eff} (1-\epsilon)}{1+(\omega/\omega_\textrm{eff})^2},
\end{split}
\end{equation}
where $\omega_\textrm{eff}$ is the characteristic frequency for vortex oscillations, which is linked to the depinning frequency $\omega_\textrm{p}$ and the creep parameter $\epsilon$.
$\rho_\textrm{ff}$ is the flux-flow resistivity, the effective resistivity in the high frequency limit ($\omega \gg \omega_\textrm{p}$) where vortices flow freely,
\begin{equation}\label{eq:resCCff1}
\rho_\textrm{ff}(x,y,H_\perp) = \frac{\Phi_0}{\eta} B_\perp(x,y,H_\perp),
\end{equation}
where $\Phi_0$ is the magnetic flux quantum,
$B_\perp$ is the magnetic field perpendicular to the film inside the superconductor carried by vortices,
and $\eta$ is the viscous drag coefficient per unit vortex length associated with vortex motion.
Here, the spatial distribution of vortices is given by $B_\perp(x,y,H_\perp)$.

In the temperature range we are interested in, $\lesssim\,$100 mK, $\epsilon \rightarrow 0$ and $\omega_\textrm{eff} \rightarrow \omega_\textrm{p}$ (see Sec.~S4 for justification);
$\omega_\textrm{p}$ and $\eta$ completely describe the complex resistivity from vortex motion.
As $\omega_\textrm{p}$ is given by $k_\textrm{p}/\eta$, where $k_\textrm{p}$ is the restoring force constant of a pinning potential per unit vortex length, the loss parameters associated with vortex motion are $\eta$ and $k_\textrm{p}$.

A number of studies on the time-dependent GL equations showed that there are two different mechanisms for $\eta$: Tinkham mechanism and Bardeen-Stephen mechanism.\cite{tinkham, kopnin}
If the material is an extreme type-II and the magnetic field is well below $H_\textrm{c2}$ ($B_\perp \ll \mu_0 H_\textrm{c2}^\perp$), both mechanisms have the form
\begin{equation}\label{eq:resCCff2}
\rho_\textrm{ff} \approx \alpha_\textrm{vm} \rho_\textrm{n} \frac{B_\perp}{\mu_0 H_\textrm{c2}^\perp},
\end{equation}
where $\alpha_\textrm{vm}$ is a constant of order unity.
By comparing Eqs.~\eqref{eq:resCCff1} and \eqref{eq:resCCff2}, we find
\begin{equation}\label{eq:resCCeta}
\eta \approx \frac{\mu_0 H_\textrm{c2}^\perp \Phi_0}{\alpha_\textrm{vm}\rho_\textrm{n}}.
\end{equation}

A crucial property is that, according to Eq.~\eqref{eq:resCCvm}, the dependence of $\rho_\textrm{vm,1}$ and $\rho_\textrm{vm,2}$ on $B_\perp$ is the same.
Hence if the field dependence of $\rho_1$ and $\rho_2$ is qualitatively different, it implies that quasiparticles, $\rho_\textrm{tf,2}$, are the major contributors to $\rho_2$.
Note that also the loss contribution induced by excitations in a vortex core is a quasiparticle contribution.

To fully understand the microwave loss, we need to know the vortex distribution.
The vortex distribution is determined by the cooling history and the pinning strength.
For a sample cooled in a magnetic field (field cooling), vortices are homogeneously distributed regardless of the pinning strength, i.e., $B_\perp$ in Eq.~\eqref{eq:resCCff1} becomes a constant.
As a result, $\eta$ and $k_\textrm{p}$ can be obtained in a straightforward way.\cite{song2009a}

For a sample cooled without a magnetic field (zero-field cooling), followed by turning on a magnetic field, we consider two extreme cases.
If the critical current density associated with vortex pinning is much lower than the depairing current density (weak pinning limit), a high surface barrier exists and vortices accumulate near the center of the superconductor, called a vortex dome, due to the strong repulsive interaction between the Meissner current along edges and vortices.\cite{brandt2013, schuster1994a, zeldov1994b, zeldov1994c, schuster1994b, maksimov1995, willa2014}
If the critical current density is comparable to the depairing current density (strong pinning limit), then vortices accumulate near the edges of the sample (Bean-type model).\cite{schuster1994b, brandt1993b, zeldov1994a}
In this limit, the surface barrier is strongly suppressed by the pinning potentials.\cite{kuznetsov1999}
For our resonators, the Bean--Livingston barrier is the dominant surface barrier. The geometrical barrier is unimportant because the film thickness $d$ is small enough to satisfy $d \lesssim \lambda \ll W$, where $W$ is the width of the strips (see Tables~\ref{tab:device} and \ref{tab:fit}).

\section{Methods}
\label{sec:method}

\begin{figure}
\centering
\includegraphics[scale=0.5]{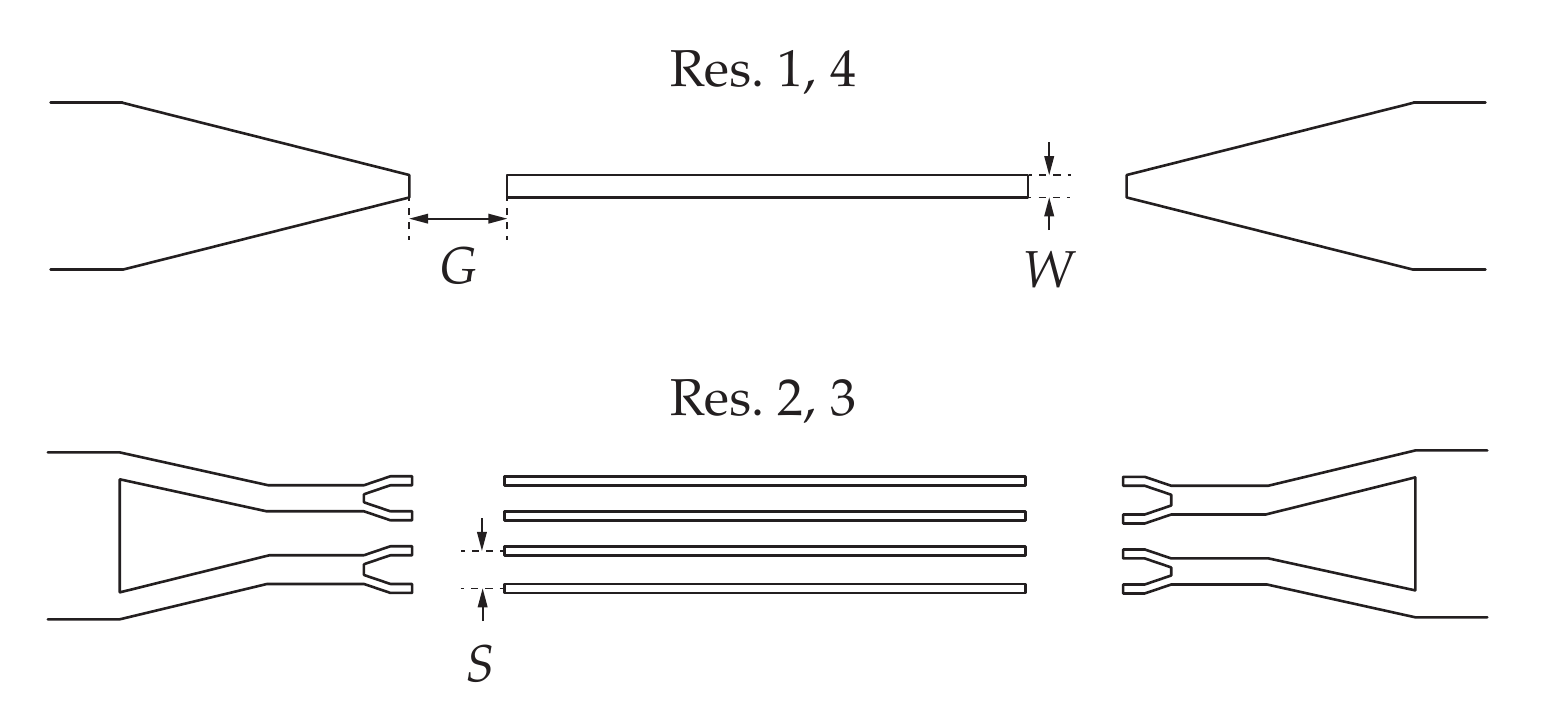}
\caption{\label{fig:config} Geometry of the resonators. Resonators 1 and 4 are single-strip resonators; resonators 2 and 3 are four-strip resonators.\cite{mohebbi2014}
$G$ is the gap between the feedline and the resonator. $W$ is the width of a strip. $S$ is the spacing between center of strips.
The values of $G$, $W$, and $S$ are given in Table~\ref{tab:device}.}
\end{figure}

\begin{table*}
\caption{Properties of thin films. $d$ is the film thickness. $T_\textrm{c}$ is the critical temperature. Residual resistivity ratio (RRR) is defined by $R(\textrm{300 K})/R(T_\textrm{c})$.}
\label{tab:film}\centering
\begin{ruledtabular}
\begin{tabular}{c c c c c r c c c c c}
\noalign{\smallskip}
	&	\multicolumn{5}{c}{Strip}	&	\multicolumn{5}{c}{Ground plane} \\
\noalign{\smallskip}
\cline{2-6} \cline{7-11}
\noalign{\smallskip} \noalign{\smallskip}
Wafer	&	$d$	&	Orient.	&	$T_\textrm{c}$		&	$\rho_\textrm{n}$	&	RRR	&	$d$	&	Orient.	&	$T_\textrm{c}$		&	$\rho_\textrm{n}$	&	RRR \\
&	(nm)	&		&	(K)		&	($\mu\Omega \cdot$cm)	&		&	(nm)	&		&	(K)		&	($\mu\Omega \cdot$cm)	&	 \\
\noalign{\smallskip} \hline \noalign{\smallskip} \noalign{\smallskip}
A	&	50.5	&	(111)	&	9.30		&	\hphantom{1}2.9	&	6.3\hphantom{.}		&	48.4	&	(110), (111)	&	8.75	&	3.9	&	4.9	 \\
B	&	98.9	&	(111)	&	9.50		&	\hphantom{1}1.1	&	15.2\hphantom{.}	&	96.5	&	(110)	&	9.23	&	3.1	&	6.2	 \\
C	&	50\hphantom{.0}	&	(110)	&	7.2\hphantom{0} 	&	17\hphantom{.0}	&	1.7\hphantom{.}	&	50\hphantom{.0}	&	(110)	&	&	& \\
\end{tabular}
\end{ruledtabular}
\end{table*}

\begin{table*}
\caption{Resonator information. The length of a strip is the same for all resonators, 5725 $\mu$m. $f_0$ and $Q_0$ are the resonance frequency and the loaded quality factors at $T_\textrm{MC} = 10$ mK without a magnetic field, where $T_\textrm{MC}$ is the mixing chamber temperature. $Q_\textrm{0,ex}$ and $Q_\textrm{0,in}$ are the external and the internal quality factors in the same condition, respectively. Remark summarizes film quality, film thickness, and strip width of each resonator.}
\label{tab:device}\centering
\begin{ruledtabular}
\begin{tabular}{c c c c c c c c c c}
\noalign{\smallskip}
Res.	&	Wafer		&	$W$	&	$S$	&	$G$	&	Remark	&	$f_0$	&	$Q_0$	&	$Q_\textrm{0,ex}$	&	$Q_\textrm{0,in}$	 \\
&	&	($\mu$m)	&	($\mu$m)	&	($\mu$m)	&	&	(GHz)	&	&	&	 \\
\noalign{\smallskip} \hline \noalign{\smallskip}
1	&	A	&	60	&		&	300	&	clean and thin film, wide width	&	10.0078	&	$3.34 \times 10^4$	&	$6 \times 10^4$	&	$8 \times 10^4$ \\
2	&	A	&	15	&	75	&	400	&	clean and thin film, narrow width	&	10.0792	&	$2.75 \times 10^4$	&	$1 \times 10^5$	&	$4 \times 10^4$ \\
3	&	C	&	15	&	75	&	350	&	dirty and thin film, narrow width	&	10.0728	&	$1.43 \times 10^4$	&	$4 \times 10^4$	&	$2 \times 10^4$ \\
4	&	B	&	60	&		&	300	&	clean and thick film, wide width	&	10.0255	&	$3.32 \times 10^4$	&	$6 \times 10^4$	&	$8 \times 10^4$ \\
\end{tabular}
\end{ruledtabular}
\end{table*}

\begin{figure}
\centering
\includegraphics[scale=0.5]{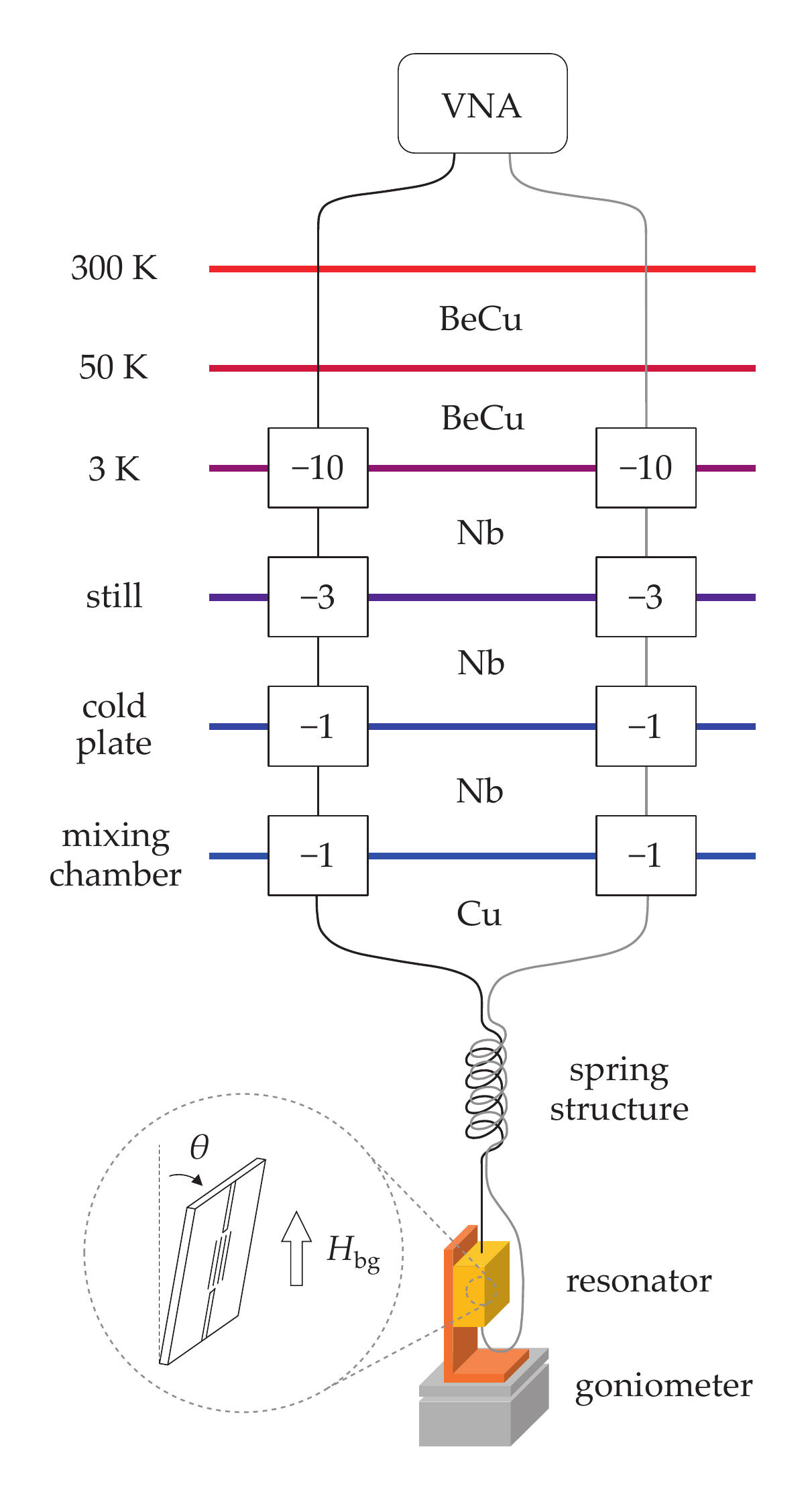}
\caption{\label{fig:cabling} Cabling and resonator configuration in the dilution refrigerator.
Coaxial compositions are shown between stages.
Numbers in boxes indicate attenuation.
A nonmagnetic chip resistor (not shown) is attached on the backside of the L-shaped bracket to apply heat pulses to the resonator.
The inset shows how a perpendicular field is applied by tilting.}
\end{figure}

Three double-side-polished 430 $\mu$m thick $2''$ diameter $c$-plane sapphire wafers were prepared and niobium films were grown by DC magnetron sputtering on both sides of the wafers.
Then the resonators were fabricated by optical lithography and dry etching.
(Details on the film growth and characterization are described in Sec.~S5.)
Table~\ref{tab:film} summarizes the basic properties of films.
The relation between $T_\textrm{c}$ and $\rho_\textrm{n}$ is similar to that reported in Refs.~\onlinecite{gubin2005, lemberger2007}.

For this study, we chose a microstrip design made of straight half-wavelength resonators, as shown in Fig.~\ref{fig:config}, without any additional structures, such as antidots or slots.
Res.~2 and 3 are multi-strip resonators.
The working principle and performance of the multi-strip resonators can be found in Ref.~\onlinecite{mohebbi2014}.
The dimensions and basic microwave properties of the resonators are shown in Table~\ref{tab:device}.

Microwave measurements were performed in a dilution refrigerator (Leiden CF250).
Schematic experimental configuration and cabling are shown in Fig.~\ref{fig:cabling}.
Resonance frequency and quality factor were measured as a function of magnetic field by collecting full $S$-parameters using a vector network analyzer (Agilent N5230A).
The resonance frequency $f_\textrm{res}$ and the loaded quality factor $Q_\textrm{load}$ were obtained by fitting the magnitude of the measured $S_{21}$ to a complex Lorentzian function as follows:
\begin{equation}\label{eq:comLor}
\left|S_{21}(f)\right| 
= \left|\frac{S_\textrm{21,max}}{1 + \textrm{i}2Q_\textrm{load}\left(\dfrac{f}{f_\textrm{res}} - 1 \right)} + A\textrm{e}^{\textrm{i}\phi} \right|,
\end{equation}
where $f$ is the excitation frequency, and $S_\textrm{21,max}$ is the maximum of the transmission coefficient.
In Eq.~\eqref{eq:comLor}, the second term is a complex background due to the direct coupling between the input and output ports through radiation.\cite{sage2011}
The fitting parameters are $S_\textrm{21,max}$, $f_\textrm{res}$, $Q_\textrm{load}$, $A$, and $\phi$.

The external quality factor $Q_\textrm{ex}$ was obtained using the formula $Q_\textrm{load} = Q_\textrm{ex} 10^{-\textrm{IL}/20}$, where IL is the insertion loss in dB.\cite{sage2011}
The insertion loss was estimated by subtracting the losses between the vector network analyzer and the package from $S_\textrm{21,max}$ in dB.
The internal quality factor $Q_\textrm{in}$ was obtained from the relation $Q_\textrm{load}^{-1} = Q_\textrm{ex}^{-1} + Q_\textrm{in}^{-1}$.

The circulating power $P_\textrm{circ}$ was kept at about $-20$ dBm throughout the measurements. This value was high enough to suppress the loss due to two-level systems in the resonator dielectrics;\cite{sage2011}
this value was also roughly 20 dB lower than the power where the quality factor is suppressed due to the nonlinearity.
$P_\textrm{circ}$ was estimated using $P_\textrm{circ} = \pi^{-1} P_\textrm{inc} Q_\textrm{load} 10^{-\textrm{IL}/20}$, where $P_\textrm{inc}$ is the incident power on the input capacitor of the resonator.\cite{sage2011}

Two cooling procedures were used: zero-field cooling (ZFC) and heat-pulsing (HP).
For HP, a heat pulse (0.16 W for 5 s) is applied to completely suppress superconductivity.
The resonator was then cooled in field.
Note that a heat pulse was applied for each magnetic field value, unlike the ordinary field-cooling procedure in many studies.

To apply a perpendicular field, the resonator is tilted in a background field parallel to the microwave current $H_\textrm{bg}$ by up to $\pm 3$ deg using a goniometer (Attocube ANGt101), as shown in the inset of Fig.~\ref{fig:cabling}.
$H_\perp$ is obtained by $H_\perp = H_\textrm{bg} \sin\theta$.
The precision at 100 mK is roughly $\pm 5$ mdeg.
Initial alignment was done at $\mu_0 H_\textrm{bg} = 0.7$ T after the HP procedure.
The position of the goniometer that maximized $f_\textrm{res}$ and $Q_\textrm{load}$ was assumed to be $\theta = 0^\circ$.
A spring structure is employed between the resonator and the mixing chamber to make the cables flexible.

For the rest of this paper, we will write the resonance frequency as $f$ and the loaded quality factor as $Q$ for simplicity.

\section{Results and Analysis}
\label{sec:result}

\begin{figure*}
\centering
\includegraphics[scale=0.5]{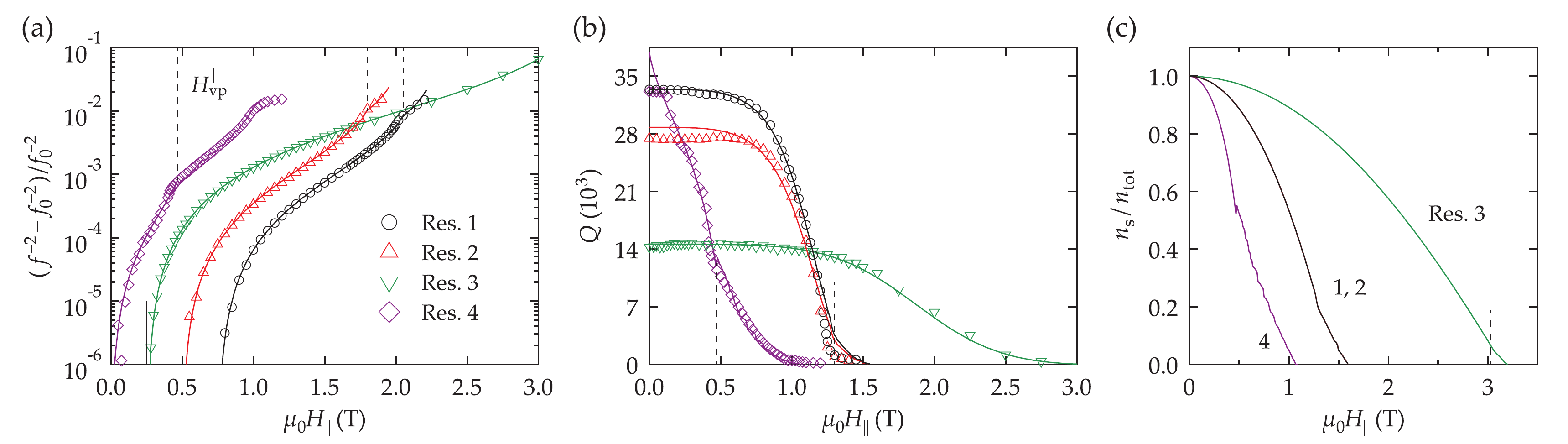}
\caption{\label{fig:paraField} (a,b) Parallel magnetic field $H_\parallel$ dependence of $f^{-2}$ and $Q$ after ZFC.
Solid lines are from calculations with parameters in Table~\ref{tab:fit}.
In (a), $f^{-2}$ data are shifted by 0.25 T steps for clarity. The offsets are indicated by vertical solid lines.
(c) Parallel field dependence of $n_\textrm{s}$ calculated by the GL equations.
Vertical dashed lines indicate the vortex penetration field parallel to the film $H_\textrm{vp}^\parallel$ obtained from the solution of the GL equations.
$H_\textrm{vp}^\parallel$ of Res.~3 is shown in (c) because its value is out of the experimental range.
We also note that $H_\textrm{vp}^\parallel$ of Res.~1 and 2 are identical, even though they don't appear so in (a) due to the visual offset mentioned above.
$T_\textrm{MC}$ is less than 20 mK.
}
\end{figure*}

\begin{figure*}
\centering
\includegraphics[scale=0.5]{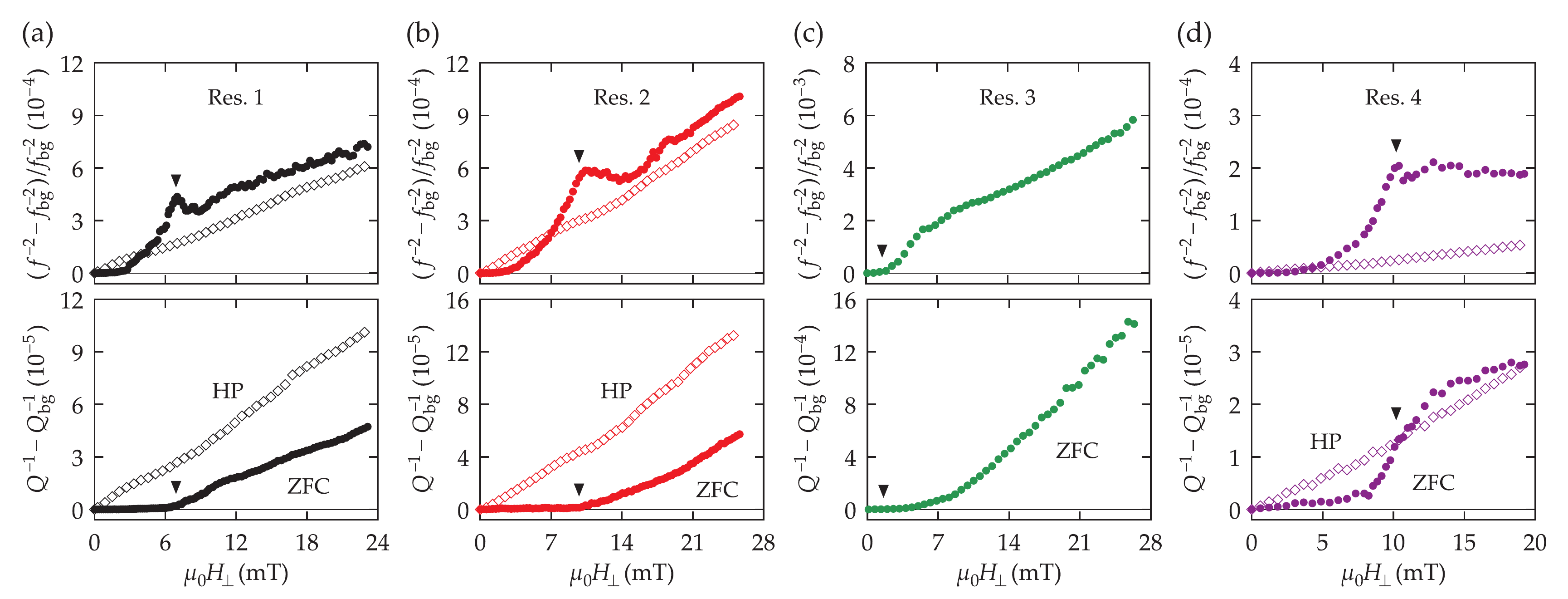}
\caption{\label{fig:perpField} Perpendicular magnetic field $H_\perp$ dependence of $f^{-2}$ and $Q^{-1}$.
The subscript ``bg'' means that the quantity is measured with $H_\textrm{bg}$, but without tilting: $f_\textrm{bg} \equiv f(H_\parallel=H_\textrm{bg}, \theta=0)$ and $Q_\textrm{bg} \equiv Q(H_\textrm{bg}, 0)$.
$T_\textrm{MC} \approx 100$ mK and $\mu_0 H_\textrm{bg} = 0.35$ T.
The cooling procedure for Res.~1, 2, and 4 is both ZFC and HP; for Res.~3, ZFC only.
Note that the scales for $f^{-2}$ and $Q^{-1}$ are different from resonator to resonator.
Arrows indicate the crossover field $H_\textrm{X}$, which is determined by plots of $Q$ vs. $f^{-2}$ (Fig.~\ref{fig:perpField_Hbg}).
}
\end{figure*}

\begin{table*}
\caption{Loss parameters extracted from Figs.~\ref{fig:paraField} and \ref{fig:perpField}.
Two characteristic fields, related to switching of the dominant loss mechanism, are also shown.
$Q_\textrm{0,fit}$ is the zero-field $Q$ determined by fitting;
$\rho_\textrm{n,fit}$ is the residual resistivity obtained from fitting;
$H_\textrm{vp}^\perp$ is the vortex penetration field perpendicular to the film (see Sec.~\ref{sec:freqAnomaly});
$H_\textrm{X}$ is the crossover field (see Sec.~\ref{sec:qfPlot}).
Experimental conditions for each group of loss parameters are indicated in parentheses.
For Res.~3, the ZFC data above 14 mT were used to obtain the loss parameters associated with vortex motion because of the absence of the HP data.}
\label{tab:fit}\centering
\begin{ruledtabular}
\begin{tabular}{c c c c c c c c c c c c}
\noalign{\smallskip}
& \multicolumn{6}{c}{Quasiparticle generation} & \multicolumn{3}{c}{Vortex motion} & \multicolumn{2}{c}{Char. field} \\
& \multicolumn{6}{c}{($H_\parallel$, ZFC)} & \multicolumn{3}{c}{($H_\perp$, HP)} & \multicolumn{2}{c}{($H_\perp$, ZFC)} \\
\noalign{\smallskip} \cline{2-7} \cline{8-10} \cline{11-12} \noalign{\smallskip}
Res.	&	$\lambda_\textrm{0}$	&	$\kappa$	&	$\mu_0 H_\textrm{c}$	&	$Q_\textrm{0,fit}$	&	$\beta$	&	$\rho_\textrm{n,fit}$	&	$\omega_\textrm{p}/2\pi$	&	$\eta$	&	$k_\textrm{p}$	&	$\mu_0 H_\textrm{vp}^\perp$	&	$\mu_0H_\textrm{X}$	\\
&	(nm)	&		&	(mT)	&		&		&	($\mu\Omega \cdot$cm)	&	(GHz)	&	(N$\cdot$s/m$^2$)	&	(N/m$^2$)	&	(mT)	&	(mT)		\\
\noalign{\smallskip} \hline \noalign{\smallskip} \noalign{\smallskip}
1	&	\hphantom{1}52	&	2.43	&	270	&	$3.34 \times 10^4$	&	1.8	&	\hphantom{1}2.9	&	58	&	$1.3 \times 10^{-7}$	&	$4.8 \times 10^{4}$	&	\hphantom{1}6.9	&	\hphantom{1}6.9 \\
2	&	\hphantom{1}52	&	2.43	&	270	&	$2.88 \times 10^4$	&	1.8	&	\hphantom{1}2.9	&	62	&	$1.2 \times 10^{-7}$	&	$4.7 \times 10^{4}$	&	10.4			&	\hphantom{1}9.8 \\
3	&	162		&	6.5\hphantom{0}	&	190	&	$1.46 \times 10^4$	&	2.2	&	17\hphantom{.0}	&	24	&	$3.6 \times 10^{-8}$	&	$5.9 \times 10^{3}$	&	\hphantom{1}5.5	&	\hphantom{1}2.2 \\
4	&	\hphantom{1}43	&	1.80	&	250	&	$3.80 \times 10^4$	&	0.4	&	\hphantom{1}0.6	&	20	&	$1.9 \times 10^{-6}$	&	$2.3 \times 10^{5}$	&	10.2			&	10.2 \\
\end{tabular}
\end{ruledtabular}
\end{table*}

\subsection{Loss Parameters}
\label{sec:lossPara}


In a magnetic field parallel to the film, the resonator performance is expected to be governed by quasiparticle generation due to the absence of the Lorentz force on vortices.
Hence the loss parameters for quasiparticle generation can be obtained from the $H_\parallel$ dependence of $f$ and $Q$ as shown in Fig.~\ref{fig:paraField}.
The loss parameters for calculated curves are shown in Table~\ref{tab:fit}.
The parameters $\lambda_0$, $\kappa$, and $H_\textrm{c}$ were mostly determined by comparing the observed and expected $f^{-2}$.
The expected $f^{-2}$ was calculated using Eqs.~\eqref{eq:magInduc}--\eqref{eq:con2TF} and $n_\textrm{s}(H_\parallel)$ in Fig.~\ref{fig:paraField}(c).
Here, $n_\textrm{s}(H_\parallel)$ was obtained by solving the time-dependent GL equations.
Then, $Q_\textrm{0,fit}$, $\beta$, and $\rho_\textrm{n,fit}$ were obtained from $Q$ via similar procedures.
(For details, see Sec.~S3.)

The measured data and calculated curves agree well.
In Table~\ref{tab:fit}, one can see that the dirtier film shows longer $\lambda_0$, higher $\kappa$, and lower $H_\textrm{c}$, as expected.\cite{tinkham, matsushita}
The values of $\lambda_0$, $\kappa$, and $H_\textrm{c}$ are reasonable, compared to previous reports.\cite{gubin2005, lemberger2007, koch1974, halbritter2005}
These results support that quasiparticle generation is the dominant loss mechanism in a parallel field and the loss can be understood quantitatively using the two-fluid model incorporated with the time-dependent GL equations.

Solving the GL equations for the actual geometry of resonators enabled us to determine the vortex penetration field parallel to the film $H_\textrm{vp}^\parallel$ of each resonator, indicated by vertical dashed lines in Fig.~\ref{fig:paraField}.
Note that the $H_\parallel$ dependence of $n_\textrm{s}$ changes from quadratic to linear above $H_\textrm{vp}^\parallel$.
This results in a change in the slope of $f^{-2}(H_\parallel)$ and $Q(H_\parallel)$ at 1.3 T (Res.~1 and 2) and 0.47 T (Res.~4).

In Table~\ref{tab:fit}, $\beta$ varies between 0.4 and 2.2, and the film with higher $d/\lambda_0$ shows lower $\beta$.
The microscopic description of these results using the standard theoretical expressions of the complex conductivity\cite{mattis1958, zimmermann1991, dressel} seems to be challenging,\cite{white1964} because we cannot equate the order parameter, which is obtained from $f^{-2}$, to the energy gap in the presence of a perturbation that breaks the time-reversal symmetry.\cite{maki}
Although the energy gap suppression with field itself is understood well,\cite{maki, usadel1970, anthore2003} we know of no well-established expression that converts the order parameter to the energy gap for a type-II superconducting thin film in a parallel field.
In our case, such an expression also needs to consider the spatial variation of the order parameter to account for vortices and the $d/\lambda_0$ dependence of $\beta$.


Figure~\ref{fig:paraField}(b) and Tables~\ref{tab:device} and \ref{tab:fit} show the trade-off between $Q_\textrm{0,in}$ and robustness against $H_\parallel$.
The optimal condition for balancing these two factors can be written in terms of $d/\lambda_0$.
For $d/\lambda_0 < 1$ (Res.~3), $Q$ does not change much by $H_\parallel$ up to 1 T, but $Q_\textrm{0,in}$ is low;
for $d/\lambda_0 > 1$ (Res.~4), $Q_\textrm{0,in}$ is high, but $Q$ drops quickly in $H_\parallel$.
Therefore, the resonator satisfying $d/\lambda_0 \approx 1$ (Res.~1 and 2) is the best choice for X-band ESR applications, which require a magnetic field of 0.35 T.


Loss parameters associated with vortex motion are obtained by the HP procedure due to the homogeneous vortex distribution, as mentioned in Sec.~\ref{sec:vm}.
(For details, see Sec.~S4.)
Figure~\ref{fig:perpField} shows the $H_\perp$ dependence of $f^{-2}$ and $Q^{-1}$ after HP.
For all resonators, $f^{-2}$ and $Q^{-1}$ vary linearly with the field and the intercept on the $H_\perp$ axis is almost zero.
These results indicate continuous occupation of vortices and a very low lower critical field.\cite{stan2004, kuit2008}

Note that, in Table~\ref{tab:fit}, the better quality film shows a higher $k_\textrm{p}$.
The reason is that, if the film is too dirty, the Bardeen-Cooper-Schrieffer coherence lengths deep within the grain and in the vicinity of the grain boundary are similar. This results in the pinning mechanism becoming inefficient, although there may be more pinning sites.\cite{matsushita, zerweck1981}
The better quality film has a higher $\eta$, as expected in Eq.~\eqref{eq:resCCeta}.

\subsection{Frequency Anomaly}
\label{sec:freqAnomaly}

\subsubsection{Resonator 1}
\label{sec:res1}

A key feature of Fig.~\ref{fig:perpField} after ZFC is that an anomaly (peak/dip) appears in the $f^{-2}$ data.
This frequency anomaly is an indication of a partial release of the Meissner current along the edges, accompanied by vortex injection.
This phenomenon is due to the strong repulsive interaction between the Meissner current and vortices.\cite{fetter1980, kogan1994, geim2000, peeters2002}
The field at which the frequency anomaly occurs is the vortex penetration field perpendicular to the film $H_\textrm{vp}^\perp$, i.e., the field at which the surface barrier is fully suppressed.

To support the above statement, we note that the $H_\perp$ dependence of $f^{-2}$ and $Q^{-1}$ is different below and above $H_\textrm{vp}^\perp$.
We start by exploring the dominant loss mechanism in both low and high field regimes of Res. 1 (Fig.~\ref{fig:perpField}(a)).

Below 7 mT, quasiparticle generation is the dominant contribution to $f^{-2}$ and $Q^{-1}$, suggesting the existence of a large surface barrier.
This is reflected in that 
(i) the $H_\perp$ dependences of $f^{-2}$ and $Q^{-1}$ are qualitatively different from each other (see Sec.~\ref{sec:vm}), and
(ii) $f^{-2}$ grows roughly quadratically.\cite{sharvin1961, douglass1961, sridhar1989, samkharadze2016, healey2008}
This is based on the approximate relation $(f^{-2} - f_0^{-2}) \propto -n_\textrm{s}$ (Eqs.~\eqref{eq:addInduc}, \eqref{eq:freqL}, and \eqref{eq:con2TF}) and
$n_\textrm{s}$ is suppressed approximately quadratically with a magnetic field (Fig.~\ref{fig:paraField}(c)).

Above 11 mT, vortex motion is the main contribution to microwave loss.
Moreover, our results imply that a significant number of vortices are pinned near the center of the strip compared to the Bean-type model, even at the early stage of vortex penetration.
The number of vortices increases linearly with $H_\perp$ if the vortices accumulate near the center of the strip.\cite{maksimov1995}
Since the microwave current density near the center is roughly homogeneous and much lower than at the edges (Fig. S1 in Sec. S2), the resulting $f^{-2}$ and $Q^{-1}$ are expected to be linear functions of $H_\perp$, and their slope should be less than that of the corresponding HP data (see Eqs.~\eqref{eq:dissE} and \eqref{eq:resCCff1}).\cite{bothner2012a}
These expectations are consistent with our results.
In addition, the existence of a large surface barrier indicates that the critical current density associated with vortex pinning is significantly lower than the depairing current density.
Hence, the vortex distribution is expected to deviate from the Bean-type model, which is valid when the critical current density is comparable to the depairing current density (Sec.~\ref{sec:vm}).

\begin{figure*}
\centering
\includegraphics[scale=0.49]{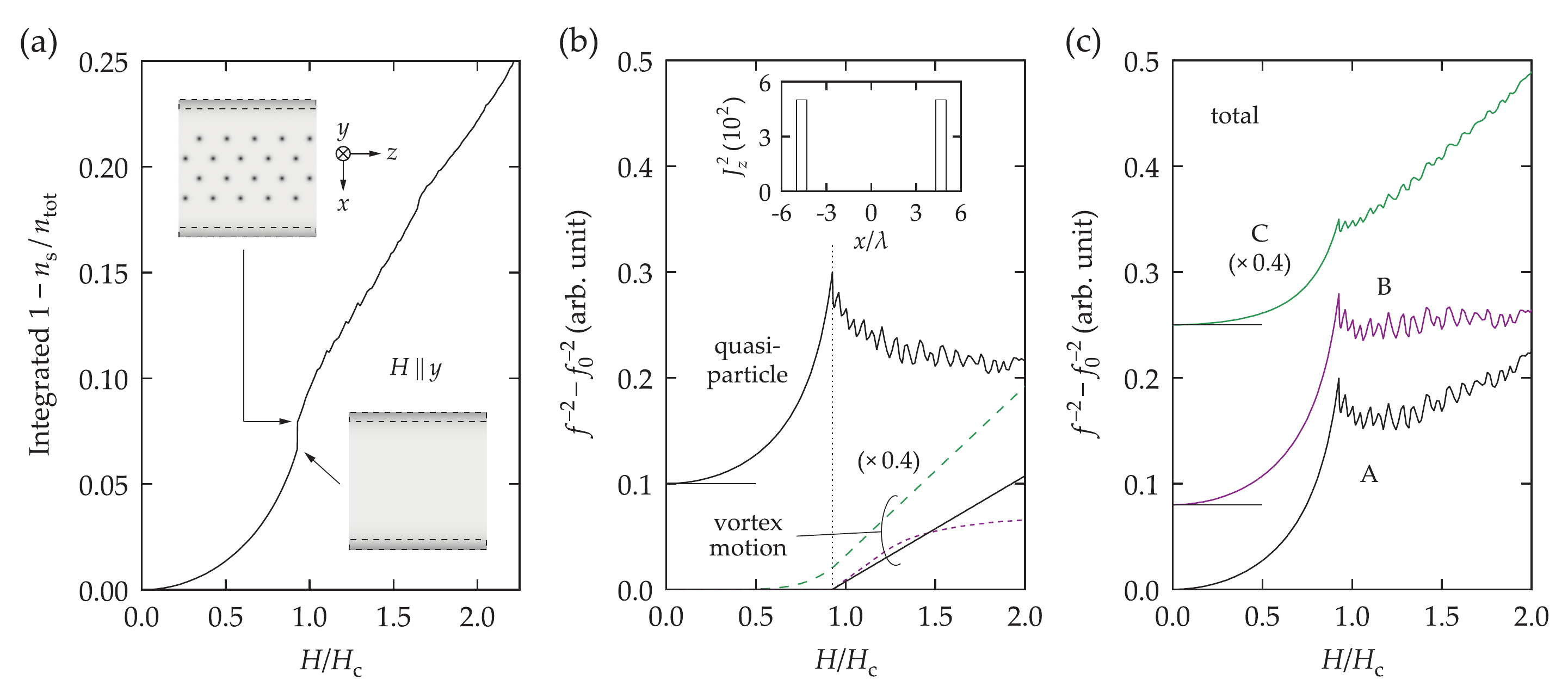}
\caption{\label{fig:GL} Simulation of $f^{-2}$ field dependence.
(a) $1 - n_\textrm{s}/n_\textrm{tot}$, integrated over the geometry as a function of $H$.
The insets show the normalized $|\psi(x,y)|$ before and after vortex penetration;
darker is lower $|\psi(x,y)|$.
Dashed rectangles are regions where the microwave current density is high.
(b) $f^{-2}$ from quasiparticle generation and vortex motion.
The quasiparticle contribution is obtained from integration of $|\psi(x,y)|^{-2}$ with the weighting function $J_z^2(x)$.
For the vortex motion contribution, curves imitating $Q^{-1}$ in Fig.~\ref{fig:perpField} are used.
The dotted line indicates the vortex penetration field.
The inset shows $J_z^2(x)$:
$J_z^2=500$ from $x = \pm4.3\lambda$ to $\pm W/2$ and $J_z^2=1$ for others.
(c) Total $f^{-2}$, which is a summation of the contributions of quasiparticle generation and vortex motion in (b).
The green lines in (b,c) are scaled with a factor of 0.4.
Some data are shifted for clarity.
The resonator-simulation correspondence is (Res.~1 and 2, A), (Res.~3, C), and (Res.~4, B).}
\end{figure*}

From the discussions so far, we see that the dominant loss mechanism switches from quasiparticle generation to vortex motion around the field where the frequency anomaly occurs.
To show how the suppression of the surface barrier and the switching of the dominant loss mechanism yield this frequency anomaly, we solved the time-dependent GL equations for a slab geometry with infinite thickness and length ($y$ and $z$ axes in the inset of Fig.~\ref{fig:GL}(a)) in a magnetic field. (For details on the simulation conditions, see Sec.~S1.)
We use this geometry because the only difference between an infinite slab and a thin film is the strength of the interaction between the Meissner current and vortices.
For a thin film, the interactions are stronger because the Meissner current and vortices interact mostly via stray magnetic fields (long-range interaction), while for a slab the interactions are exponential screening (short-range);\cite{pearl1964, fetter1980, kogan1994}
the essential physics will remain intact.

Figure~\ref{fig:GL}(a) shows how $n_\textrm{s}$ decreases with a magnetic field.
The jump at the magnetic field $H \approx 0.93 H_\textrm{c}$ indicates vortex penetration.
To obtain the quasiparticle contribution to $f^{-2}$, a weighting function $J_z^2(x)$ corresponding to the microwave current density distribution is needed.
Since the geometry in (a) is neither a resonator nor a transmission line, $J_z^2(x)$ cannot be determined via the procedure described in Sec.~S2.
Here, we used a simple stepwise function as $J_z^2(x)$ (the inset of Fig.~\ref{fig:GL}(b)).
This weighting function maximizes contributions from the region within roughly $\lambda$ of the edges (shown by dashed lines in the insets of Fig.~\ref{fig:GL}(a)), and the ratio between the values at the edge and the center is similar to that of our resonators.
(See Sec.~S2 for the ratio of $|J_z|^2$ between the edge and the center of the strips.)
The result is shown in Fig.~\ref{fig:GL}(b).
$f^{-2}$ shows a peak at the vortex penetration field, indicating that the number of quasiparticles near the edges are reduced, i.e., the Meissner current is partly released,
while the total number of quasiparticles (including vortex cores) of the entire sample increases.

The vortex motion contribution is also shown in Fig.~\ref{fig:GL}(b).
As the width of the geometry for the simulation is much more confined than for the real strips and it does not have any pinning sites, curves imitating $Q^{-1}$ in Fig.~\ref{fig:perpField} are used as the vortex motion contribution.
The total $f^{-2}$ in Fig.~\ref{fig:GL}(c) can be obtained by adding these vortex motion contributions to the quasiparticle contribution.
The resulting curve A is quite similar to $f^{-2}$ of Res.~1 (Fig.~\ref{fig:perpField}(a)).

The oscillatory behavior in Fig.~\ref{fig:GL}(b,c) after vortex penetration is due to oscillation of the Meissner current during the subsequent injection of vortices;
hence it can be understood as small variations of the frequency anomaly.
These oscillations were experimentally observed as shown in Fig.~\ref{fig:perpField} above $H_\textrm{vp}^\perp$.

Taken together, the inhomogeneous microwave current density distribution enables us to see the partial release of the Meissner current, which has previously only been observed in mesoscopic systems.\cite{geim2000, peeters2002}

\subsubsection{Other Resonators}
\label{sec:otherRes}

The frequency anomaly of Res.~2 appears at a higher field (Fig.~\ref{fig:perpField}(b)).
The reason is that, if the strip is narrower, less field accumulates at the edges.
$H_\textrm{vp}^\perp$ is expected to be proportional to $1/\sqrt{W}$,\cite{kuznetsov1999, brandt2013} suggesting that $H_\textrm{vp}^\perp$ of Res.~2 would be twice as high as that of Res.~1.
The actual value is somewhat less (Table~\ref{tab:fit}), likely due to edge imperfections.
Here, the field accumulation due to the multi-strip geometry is expected to be small because the distance between the strips is five times larger than the strip width.\cite{willa2014}

The frequency anomaly of Res.~3 (Fig.~\ref{fig:perpField}(c)) is weak compared to others.
This is due to low $\eta$ and $\omega_\textrm{p}$ (Table~\ref{tab:fit}), resulting in a large contribution from vortex motion.
The curve C in Fig.~\ref{fig:GL}(c) shows such a case.

For Res.~4 (Fig.~\ref{fig:perpField}(d)), $f^{-2}$ does not change significantly between 10 and 20 mT.
This reflects that the vortex motion contribution is weaker than other resonators
(Res.~4 has the highest $\eta$ and the lowest $\omega_\textrm{p}$)
and the vortex distribution is close to the Bean-type model as a consequence of the stronger pinning as mentioned in Sec.~\ref{sec:vm}.
Since the microwave current density is high near the edge, the initial change in $f^{-2}$ due to the vortex injection is large;
as the field increases, the slope of $f^{-2}$ decreases.\cite{bothner2012b}
This is represented by the curve B in Fig.~\ref{fig:GL}(c).

By comparing Res. 2 and 3, we can study how $H_\textrm{vp}^\perp$ is affected by the film quality.
For the geometries of our resonators, the depairing current density scales with $1/\Lambda$,\cite{clem2011} while the Meissner current density scales with $1/\sqrt{\Lambda}$,\cite{kuznetsov1999} where $\Lambda$ is the screening length given by $2\lambda \coth(d/\lambda)$.\cite{klein1990, irz1995, wei1996}
Hence, the Meissner current density of the film with longer $\Lambda$ meets the depairing current density earlier, i.e., the surface barrier is fully suppressed at a lower field.
Our results are consistent with this:
Res.~3, whose $\Lambda$ is nearly an order of magnitude longer than that of Res.~2, shows low $H_\textrm{vp}^\perp$ compared to that of Res.~2.

Increasing $d$ also enhances $H_\textrm{vp}^\perp$, because the Meissner current density is roughly proportional to $1/d$ for thin films if $\Lambda$ remains similar.\cite{kuznetsov1999}
This is why Res.~4 shows higher $H_\textrm{vp}^\perp$ than that of Res.~1. 
The film thickness, however, cannot be arbitrarily thick, because it needs to satisfy $d \approx \lambda_0$ (see Sec.~\ref{sec:lossPara}).

\subsection{$\textit{Q}$ vs. $\textit{f}^{\mathbf{-2}}$ Plot}
\label{sec:qfPlot}

\begin{figure*}
\centering
\includegraphics[scale=0.5]{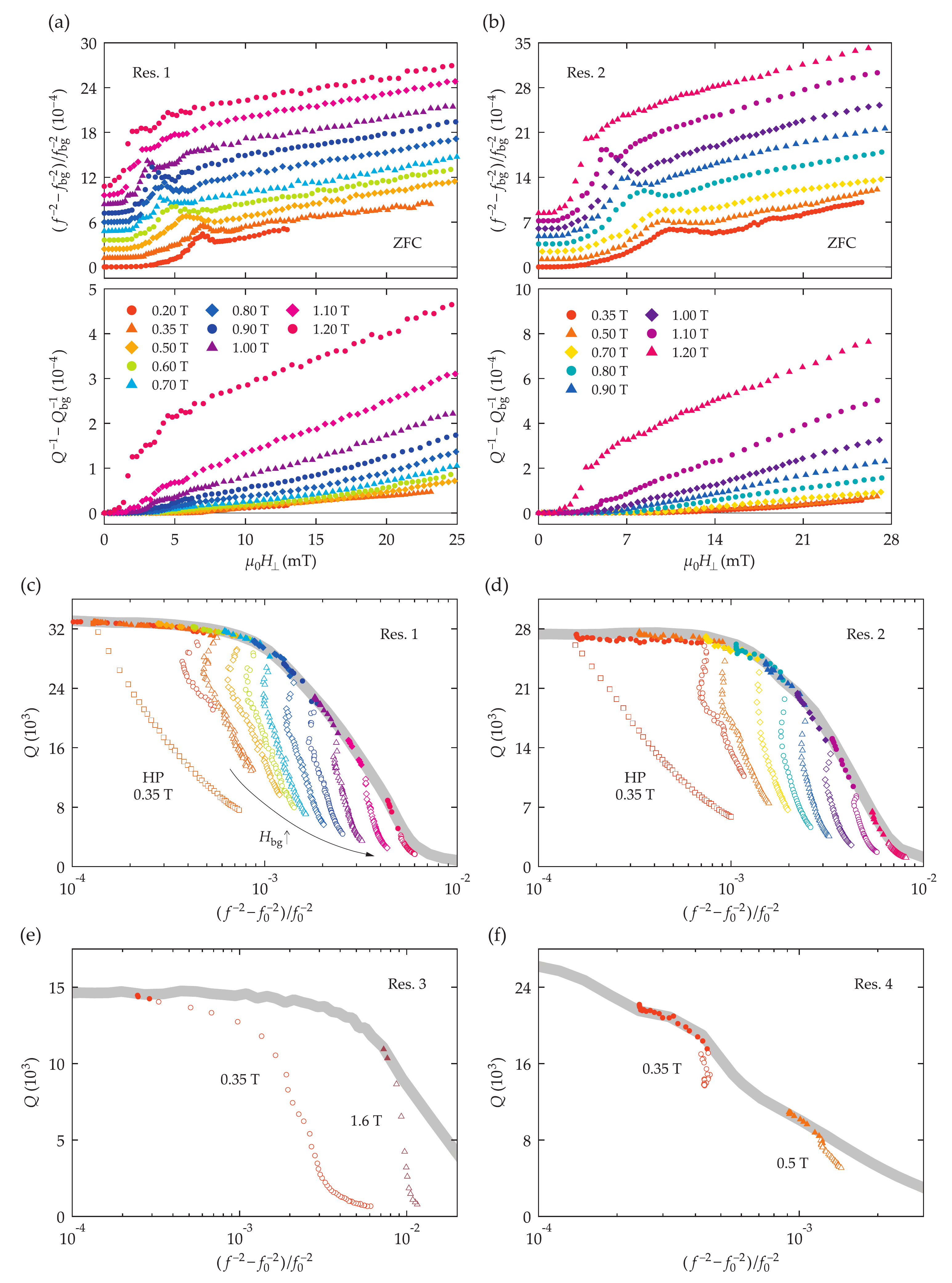}
\caption{\label{fig:perpField_Hbg} (a,b) Perpendicular field dependence of $f^{-2}$ and $Q^{-1}$ in various $H_\textrm{bg}$ for Res.~1 and 2 at $T_\textrm{MC} \approx 100$ mK.
(c--f) Plots of $Q$ vs. $f^{-2}$ for Res.~1--4.
In (c,d), the symbols and colors correspond to the same $H_\textrm{bg}$ values as in the legend from (a,b).
The gray line is the parallel field data shown in Fig.~\ref{fig:paraField}.
All data are the ZFC data, except squares in (c,d) are the HP data with $\mu_0H_\textrm{bg} = 0.35$ T.
Solid symbols are the data whose loss is dominated by quasiparticle generation;
empty symbols, dominated by vortex motion.
In (a,b), $f^{-2}$ data are shifted by $1.2 \times 10^{-4}$ steps for clarity.
}
\end{figure*}

We have discussed the dominant loss mechanism for $H_\parallel$ and $H_\perp$ with two cooling procedures.
In general, however, identifying the dominant loss mechanism is not straightforward.

Figure~\ref{fig:perpField_Hbg}(a,b) shows the $H_\perp$ dependence of $f^{-2}$ and $Q^{-1}$ in various $H_\textrm{bg}$.
The behavior of $f^{-2}$ is qualitatively similar regardless of $H_\textrm{bg}$.
One exception is the shift of the frequency anomaly to a lower $H_\perp$ for a large $H_\textrm{bg}$, which is likely due to the elongation of $\lambda$ by $H_\textrm{bg}$.
However, the behavior of $Q^{-1}$ at high $H_\textrm{bg}$ is different, especially at $\gtrsim\,$1 T:
$Q^{-1}$ increases significantly before the frequency anomaly.
Note that Res.~3 and 4 in Fig.~\ref{fig:perpField}(c,d) also show a similar behavior.

A more informative way of displaying the data is to plot $Q$ vs. $f^{-2}$,
because each loss mechanism has its own characteristic relationship between the real and the imaginary parts of the complex resistivity.
Figure~\ref{fig:perpField_Hbg}(c--f) shows this and provides a clean indication of the dominant loss mechanism.
In magnetic fields, where $Q$ follows the parallel field data (gray line), the loss is dominated by quasiparticle generation;
in fields, where $Q$ is below the gray line, the loss is dominated by vortex motion.
We identify the crossover fields $H_\textrm{X}$, where the dominant loss mechanism switches from quasiparticle generation to vortex motion, as the field where $Q$ starts to deviate from the gray line.
The arrows in Fig.~\ref{fig:perpField} are obtained through this process.

\section{Conclusion}
\label{sec:conclusion}

In this work, we have developed an approach to characterizing the magnetic field dependent microwave losses in planar superconducting resonators.
The parameters used to model the complex resistivity were obtained as the loss parameters (Table~\ref{tab:fit}) by comparing the experimentally determined $f$ and $Q$ as a function of magnetic field to calculated $f$ and $Q$.

We found that quasiparticle generation is the dominant loss mechanism for parallel magnetic fields.
For perpendicular magnetic fields, the dominant loss mechanism depends on the cooling procedure.
After HP, vortex motion is the dominant loss mechanism, while the dominating loss mechanism switches from quasiparticle generation to vortex motion after ZFC. 
For an arbitrary magnetic field direction and cooling history, the dominant loss mechanism can be readily identified from a plot of $Q$ vs. $f^{-2}$.

A frequency anomaly was observed and interpreted as partial release of the Meissner current at the vortex penetration field.
Simulations showed that the time-dependent GL equations and inhomogeneous microwave current density distribution provide an explanation of this frequency anomaly.
This suggests that the interaction between vortices and the Meissner current near the edges is crucial for understanding the magnetic field dependence of the resonator properties.

We list three conditions that a planar resonator needs to satisfy for X-band ESR of thin films:
(i) a high quality factor in a DC magnetic field of about 0.35 T,
(ii) a highly homogeneous microwave magnetic field, and
(iii) critical coupling to the external circuit.

Regarding condition (i), we have found that a niobium microstrip resonator satisfying $d \approx \lambda_0$ is a suitable choice:
it gives reasonably high $Q_\textrm{0,in}$, while the quasiparticle loss induced by $H_\parallel$ is low enough at 0.35 T.
Improving the film quality is beneficial for reducing the loss induced by quasiparticle generation (via shorter $\lambda$, see Sec.~\ref{sec:tf}) and vortex motion (via higher $\eta$ and $H_\textrm{vp}^\perp$).
Narrowing the strip width enhances $H_\textrm{vp}^\perp$, hence decreasing the number of vortices.
Most importantly, a resonator has to be aligned precisely parallel to the magnetic field to minimize both the number of vortices and their motion. 
In this respect, we believe microstrip resonators are advantageous over coplanar waveguide resonators because there is no field accumulation between the strip and the ground plane, making them more robust against misalignment with respect to the perpendicular field.

Microstrip resonators are also advantageous for condition (ii).
For microstrip resonators, a highly homogeneous microwave field is easily achievable by employing a multi-strip design, whereas applying the multi-strip design in coplanar waveguide or lumped element resonators does not seem to be straightforward.

As for condition (iii), critically coupled resonators can be made based on our results, especially the results shown in Tables~\ref{tab:device} and \ref{tab:fit}.

\section*{Supplementary Materials}

See the supplementary materials for details regarding solving the GL equations (Sec.~S1), simulating the microwave current density distribution and the stored electromagnetic energy (Sec.~S2), extracting the loss parameters (Secs.~S3 and S4), and the film growth and characterization (Sec.~S5).

\begin{acknowledgements}
S.K. thanks to A. Mitrovic, D. Carkner, Y. Ge for technical help,
T. Matsushita, V. G. Kogan, G. P. Mikitik, E. Zeldov, R. Willa, T. W. Borneman for fruitful discussions,
and the reviewers for helpful suggestions.
This work is supported by the Canada First Research Excellence Fund, 
Canada Excellence Research Chairs (grant No. 215284), 
Natural Sciences and Engineering Research Council of Canada (grant Nos. RGPIN-418579 and RGPIN-04178), 
and
Province of Ontario.
The University of Waterloo's Quantum NanoFab was used for this work. 
This infrastructure is supported by
the Canada Foundation for Innovation, 
the Ontario Ministry of Research \& Innovation, 
Industry Canada, and 
Mike \& Ophelia Lazaridis. 
\end{acknowledgements}

\newpage
$ $
\includepdf[pages=1]{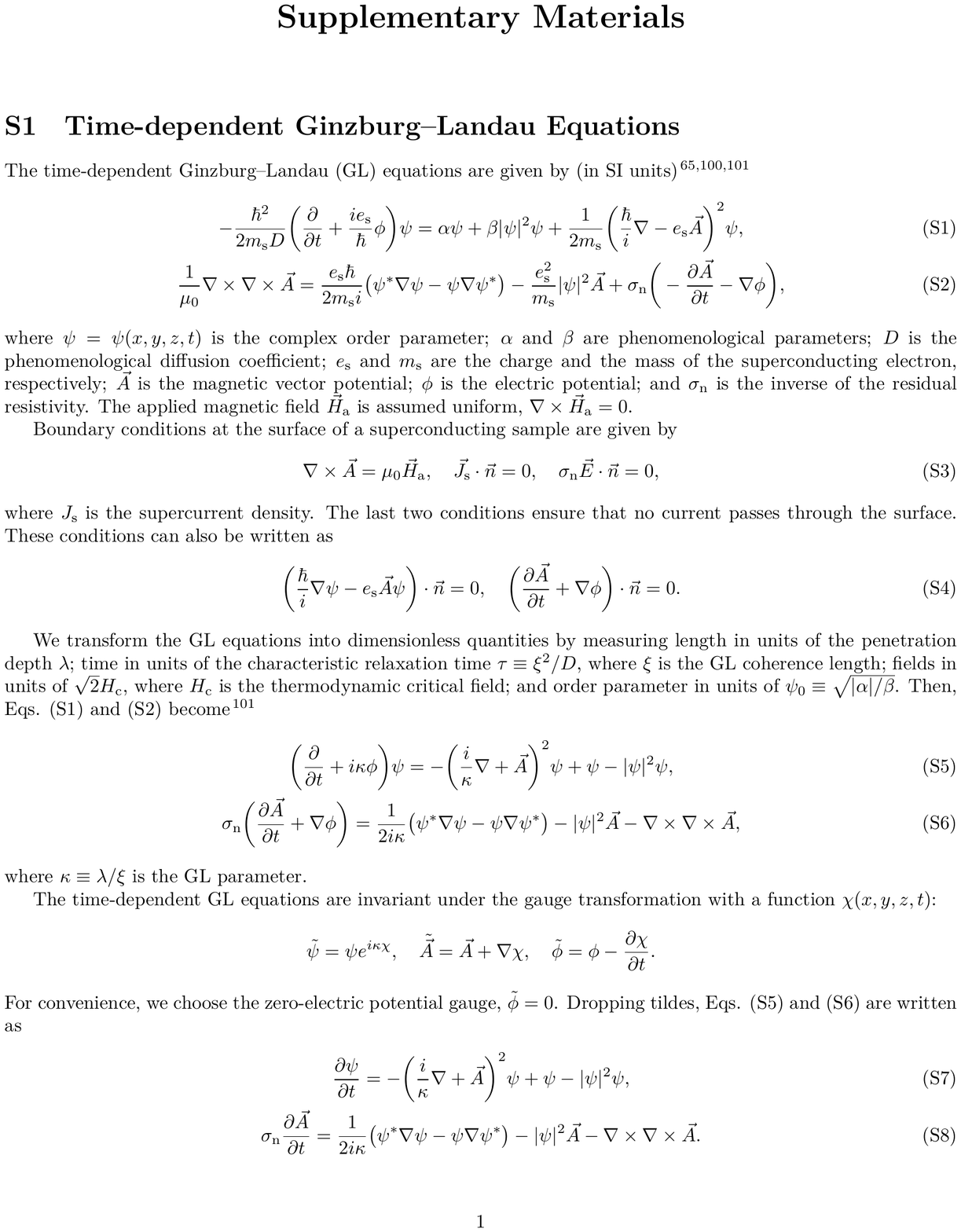}
\newpage
$ $
\includepdf[pages=2]{supp.pdf}
\newpage
$ $
\includepdf[pages=3]{supp.pdf}
\newpage
$ $
\includepdf[pages=4]{supp.pdf}
\newpage
$ $
\includepdf[pages=5]{supp.pdf}
\newpage
$ $
\includepdf[pages=6]{supp.pdf}


\begin{thebibliography}{99}


\bibitem{zmuidzinas} J. Zmuidzinas, \textit{Superconducting Microresonators: Physics and Applications}, Annu. Rev. Condens. Matter Phys. \textbf{3}, 169 (2012).
\bibitem{lancaster} M. J. Lancaster, \textit{Passive Microwave Device Applications of High-Temperature Superconductors} (Cambridge University Press, 1997).
\bibitem{hein} M. Hein, \textit{High-Temperature-Superconductor Thin Films at Microwave Frequencies} (Springer, 1999).

\bibitem{haroche} S. Haroche and J.-M. Raimond, \textit{Exploring the Quantum: Atoms, Cavities, and Photons} (Oxford University Press, 2013).
\bibitem{schoelkopf} R. J. Schoelkopf and S. M. Girvin, \textit{Wiring up Quantum Systems}, Nature \textbf{451}, 664 (2008).
\bibitem{clarke} J. Clarke and F. K. Wilhelm, \textit{Superconducting Quantum Bits}, Nature \textbf{453}, 1031 (2008).
\bibitem{gu} X. Gu, A. F. Kockum, A. Miranowicz, Y. X. Liu, and F. Nori, \textit{Microwave Photonics with Superconducting Quantum Circuits}, Phys. Rep. \textbf{718-719}, 1 (2017).

\bibitem{wallquist} M. Wallquist, K. Hammerer, P. Rabl, M. Lukin, and P. Zoller, \textit{Hybrid Quantum Devices and Quantum Engineering}, Phys. Scr. \textbf{T137}, 014001 (2009).
\bibitem{poot} M. Poot and H. S. J. van der Zant, \textit{Mechanical Systems in the Quantum Regime}, Phys. Rep. \textbf{511}, 273 (2012).
\bibitem{houck} A. A. Houck, H. E. T{\"u}reci, and J. Koch, \textit{On-Chip Quantum Simulation with Superconducting Circuits}, Nat. Phys. \textbf{8}, 292 (2012).
\bibitem{daniilidis} N. Daniilidis and H. H{\"a}ffner, \textit{Quantum Interfaces Between Atomic and Solid-State Systems}, Annu. Rev. Condens. Matter Phys. \textbf{4}, 83 (2013).
\bibitem{aspelmeyer} M. Aspelmeyer, T. J. Kippenberg, and F. Marquardt, \textit{Cavity Optomechanics}, Rev. Mod. Phys. \textbf{86}, 1391 (2014).
\bibitem{morton} J. J. L. Morton and B. W. Lovett, \textit{Hybrid Solid-State Qubits: The Powerful Role of Electron Spins}, Annu. Rev. Condens. Matter Phys. \textbf{2}, 189 (2011). 
\bibitem{xiang} Z.-L. Xiang, S. Ashhab, J. Q. You, and F. Nori, \textit{Hybrid Quantum Circuits: Superconducting Circuits Interacting with Other Quantum Systems}, Rev. Mod. Phys. \textbf{85}, 623 (2013).
\bibitem{kurizki} G. Kurizki, P. Bertet, Y. Kubo, K. M{\o}lmer, D. Petrosyan, P. Rabl, and J. Schmiedmayer, \textit{Quantum Technologies with Hybrid Systems}, Proc. Natl. Acad. Sci. U.S.A. \textbf{112}, 3866 (2015).

\bibitem{ghirri} A. Ghirri, A. Candini, and M. Affronte, \textit{Molecular Spins in the Context of Quantum Technologies}, Magnetochemistry \textbf{2017}, 3, 12 (2017).

\bibitem{staudt2012} M. U. Staudt, I.-C. Hoi, P. Krantz, M. Sandberg, M. Simoen, P. Bushev, N. Sangouard, M. Afzelius, V. S. Shumeiko, G. Johansson, P. Delsing, and C. M. Wilson, \textit{Coupling of an Erbium Spin Ensemble to a Superconducting Resonator}, J. Phys. B: At. Mol. Opt. Phys. \textbf{45}, 124019 (2012).

\bibitem{benningshof2013} O. W. B. Benningshof, H. R. Mohebbi, I. A. J. Taminiau, G. X. Miao, and D. G. Cory, \textit{Superconducting Microstrip Resonator for Pulsed ESR of Thin Films}, J. Magn. Reson. \textbf{230}, 84 (2013).
\bibitem{malissa2013} H. Malissa, D. I. Schuster, A. M. Tyryshkin, A. A. Houck, and S. A. Lyon, \textit{Superconducting Coplanar Waveguide Resonators for Low Temperature Pulsed Electron Spin Resonance Spectroscopy}, Rev. Sci. Inst. \textbf{84}, 025116 (2013).
\bibitem{probst2013} S. Probst, H. Rotzinger, S. W{\"u}nsch, P. Jung, M. Jerger, M. Siegel, A. V. Ustinov, and P. A. Bushev, \textit{Anisotropic Rare-Earth Spin Ensemble Strongly Coupled to a Superconducting Resonator}, Phys. Rev. Lett. \textbf{110}, 157001 (2013).
\bibitem{huebl2013} H. Huebl, C. W. Zollitsch, J. Lotze, F. Hocke, M. Greifenstein, A. Marx, R. Gross, and S. T. B. Goennenwein, \textit{High Cooperativity in Coupled Microwave Resonator Ferrimagnetic Insulator Hybrids}, Phys. Rev. Lett. \textbf{111}, 127003 (2013).

\bibitem{sigillito2014} A. J. Sigillito, H. Malissa, A. M. Tyryshkin, H. Riemann, N. V. Abrosimov, P. Becker, H.-J. Pohl, M. L. W. Thewalt, K. M. Itoh, J. J. L. Morton, A. A. Houck, D. I. Schuster, and S. A. Lyon, \textit{Fast, Low-Power Manipulation of Spin Ensembles in Superconducting Microresonators}, Appl. Phys. Lett. \textbf{104}, 222407 (2014).
\bibitem{grezes2014} C. Grezes, B. Julsgaard, Y. Kubo, M. Stern, T. Umeda, J. Isoya, H. Sumiya, H. Abe, S. Onoda, T. Ohshima, V. Jacques, J. Esteve, D. Vion, D. Esteve, K. M{\o}lmer, and P. Bertet, \textit{Multimode Storage and Retrieval of Microwave Fields in a Spin Ensemble}, Phys. Rev. X \textbf{4}, 021049 (2014).
\bibitem{tkalcec2014} A. Tkal{\v c}ec, S. Probst, D. Rieger, H. Rotzinger, S. W{\"u}nsch, N. Kukharchyk, A. D. Wieck, M. Siegel, A. V. Ustinov, and P. Bushev, \textit{Strong Coupling of an Er$^{3+}$-Doped YAlO$_3$ Crystal to a Superconducting Resonator}, Phys. Rev. B \textbf{90}, 075112 (2014).
\bibitem{putz2014} S. Putz, D. O. Krimer, R. Ams{\"u}ss, A. Valookaran, T. N{\"o}bauer, J. Schmiedmayer, S. Rotter, and J. Majer, \textit{Protecting a Spin Ensemble Against Decoherence in the Strong-Coupling Regime of Cavity QED}, Nature Phys. \textbf{10}, 720 (2014). 
\bibitem{wisby2014} I. Wisby, S. E. de Graaf, R. Gwilliam, A. Adamyan, S. E. Kubatkin, P. J. Meeson, A. Ya. Tzalenchuk, and T. Lindstr{\"o}m, \textit{Coupling of a Locally Implanted Rare-Earth Ion Ensemble to a Superconducting Microresonator}, Appl. Phys. Lett. \textbf{105}, 102601 (2014).

\bibitem{ghirri2015} A. Ghirri, C. Bonizzoni, D. Gerace, S. Sanna, A. Cassinese, and M. Affronte, \textit{YBa$_2$Cu$_3$O$_7$ Microwave Resonators for Strong Collective Coupling with Spin Ensembles}, Appl. Phys. Lett. \textbf{106}, 184101 (2015).
\bibitem{grezes2015} C. Grezes, B. Julsgaard, Y. Kubo, W. L. Ma, M. Stern, A. Bienfait, K. Nakamura, J. Isoya, S. Onoda, T. Ohshima, V. Jacques, D. Vion, D. Esteve, R. B. Liu, K. M{\o}lmer, and P. Bertet, \textit{Storage and Retrieval of Microwave Fields at the Single-Photon Level in a Spin Ensemble}, Phys. Rev. A \textbf{92}, 020301(R) (2015).
\bibitem{zollitsch2015} C. W. Zollitsch, K. Mueller, D. P. Franke, S. T. B. Goennenwein, M. S. Brandt, R. Gross, and H. Huebl, \textit{High Cooperativity Coupling between a Phosphorus Donor Spin Ensemble and a Superconducting Microwave Resonator}, Appl. Phys. Lett. \textbf{107}, 142105 (2015).

\bibitem{bienfait2016a} A. Bienfait, J. J. Pla, Y. Kubo, M. Stern, X. Zhou, C. C. Lo, C. D. Weis, T. Schenkel, M. L. W. Thewalt, D. Vion, D. Esteve, B. Julsgaard, K. M{\o}lmer, J. J. L. Morton, and P. Bertet, \textit{Reaching the Quantum Limit of Sensitivity in Electron Spin Resonance}, Nat. Nanotechnol. \textbf{11}, 253 (2016).
\bibitem{bienfait2016b} A. Bienfait, J. J. Pla, Y. Kubo, X. Zhou, M. Stern, C. C. Lo, C. D. Weis, T. Schenkel, D. Vion, D. Esteve, J. J. L. Morton, and P. Bertet, \textit{Controlling Spin Relaxation with a Cavity}, Nature \textbf{531}, 74 (2016).
\bibitem{bonizzoni2016} C. Bonizzoni, A. Ghirri, K. Bader, J. van Slageren, M. Perfetti, L. Sorace, Y. Lan, O. Fuhr, M. Ruben, and  M. Affronte, \textit{Coupling Molecular Spin Centers to Microwave Planar Resonators: towards Integration of Molecular Qubits in Quantum Circuits}, Dalton Trans. \textbf{45}, 16596 (2016).
\bibitem{wisby2016} I. S. Wisby, S. E. de Graaf, R. Gwilliam, A. Adamyan, S. E. Kubatkin, P. J. Meeson, A. Ya. Tzalenchuk, and T. Lindstr{\"o}m, \textit{Angle-Dependent Microresonator ESR Characterization of Locally Doped Gd$^{3+}$:Al$_2$O$_3$}, Phys. Rev. Applied \textbf{6}, 024021 (2016).

\bibitem{eichler2017} C. Eichler, A. J. Sigillito, S. A. Lyon, and J. R. Petta, \textit{Electron Spin Resonance at the Level of $10^4$ Spins Using Low Impedance Superconducting Resonators}, 
Phys. Rev. Lett. \textbf{118}, 037701 (2017).
\bibitem{astner2017} T. Astner, S. Nevlacsil, N. Peterschofsky, A. Angerer, S. Rotter, S. Putz, J. Schmiedmayer, and J. Majer, \textit{Coherent Coupling of Remote Spin Ensembles via a Cavity Bus}, Phys. Rev. Lett. \textbf{118}, 140502 (2017).


\bibitem{mohebbi2014} H. R. Mohebbi, O. W. B. Benningshof, I. A. J. Taminiau, G. X. Miao, and D. G. Cory, \textit{Composite Arrays of Superconducting Microstrip Line Resonators}, J. Appl. Phys. \textbf{115}, 094502 (2014).

\bibitem{frunzio2005} L. Frunzio, A. Wallraff, D. Schuster, J. Majer, and R. Schoelkopf, \textit{Fabrication and Characterization of Superconducting Circuit QED Devices for Quantum Computation}, IEEE Trans. Appl. Supercond. \textbf{15}, 860 (2005).

\bibitem{song2009b} C. Song, M. P. DeFeo, K. Yu, and B. L. T. Plourde, \textit{Reducing Microwave Loss in Superconducting Resonators due to Trapped Vortices}, Appl. Phys. Lett. \textbf{95}, 232501 (2009).
\bibitem{bothner2011} D. Bothner, T. Gaber, M. Kemmler, D. Koelle, and R. Kleiner, \textit{Improving the Performance of Superconducting Microwave Resonators in Magnetic Fields}, Appl. Phys. Lett. \textbf{98}, 102504 (2011).
\bibitem{bothner2012a} D. Bothner, C. Clauss, E. Koroknay, M. Kemmler, T. Gaber, M. Jetter, M. Scheffler, P. Michler, M. Dressel, D. Koelle, and R. Kleiner, \textit{Reducing Vortex Losses in Superconducting Microwave Resonators with Microsphere Patterned Antidot Arrays}, Appl. Phys. Lett. \textbf{100}, 012601 (2012).
\bibitem{bothner2012b} D. Bothner, T. Gaber, M. Kemmler, D. Koelle, R. Kleiner, S. W\"unsch, and M. Siegel, \textit{Magnetic Hysteresis Effects in Superconducting Coplanar Microwave Resonators}, Phys. Rev. B \textbf{86}, 014517 (2012).

\bibitem{deGraaf2012} S. E. de Graaf, A. V. Danilov, A. Adamyan, T. Bauch, and S. E. Kubatkin, \textit{Magnetic Field Resilient Superconducting Fractal Resonators for Coupling to Free Spins}, J. Appl. Phys. \textbf{112}, 123905 (2012).
\bibitem{deGraaf2014} S. E. de Graaf, D. Davidovikj, A. Adamyan, S. E. Kubatkin, and A. V. Danilov, \textit{Galvanically Split Superconducting Microwave Resonators for Introducing Internal Voltage Bias}, Appl. Phys. Lett. \textbf{104}, 052601 (2014).
\bibitem{singh2014} V. Singh, B. H. Schneider, S. J. Bosman, E. P. J. Merkx, and G. A. Steele, \textit{Molybdenum-Rhenium Alloy Based High-Q Superconducting Microwave Resonators}, Appl. Phys. Lett. \textbf{105}, 222601 (2014).

\bibitem{samkharadze2016} N. Samkharadze, A. Bruno, P. Scarlino, G. Zheng, D. P. DiVincenzo, L. DiCarlo, and L. M. K. Vandersypen, \textit{High-Kinetic-Inductance Superconducting Nanowire Resonators for Circuit QED in a Magnetic Field}, Phys. Rev. Applied \textbf{5}, 044004 (2016).
\bibitem{tang2016} Y.-C. Tang, H. Zhang, S. Kwon, H. R. Mohebbi, D. G. Cory, L.-C. Peng, L. Gu, H.-Z. Guo, K.-J. Jinn and G.-X. Miao, \textit{Superconducting Resonators Based on TiN/Tapering/NbN/Tapering/TiN Heterostructures}, ‎Adv. Eng. Mater. \textbf{18}, 1816 (2016).
\bibitem{ebensperger2016} N. G. Ebensperger, M. Thiemann, M. Dressel, and M. Scheffler, \textit{Superconducting Pb Stripline Resonators in Parallel Magnetic Field and Their Application for Microwave Spectroscopy}, Supercond. Sci. Technol. \textbf{29}, 115004 (2016).

\bibitem{tang2017} Y.-C. Tang,  S. Kwon, H. R. Mohebbi, D. G. Cory, and G.-X. Miao, \textit{Phonon Engineering in Proximity Enhanced Superconductor Heterostructures}, Sci. Rep. \textbf{7}, 4282 (2017).
\bibitem{bothner2017} D. Bothner, D. Wiedmaier, B. Ferdinand, R. Kleiner, and D. Koelle, \textit{Improving Superconducting Resonators in Magnetic Fields by Reduced Field Focussing and Engineered Flux Screening}, Phys. Rev. Applied \textbf{8}, 034025 (2017).


\bibitem{tinkham} M. Tinkham, \textit{Introduction to Superconductivity}, 2nd ed. (McGraw-Hill, 1996).

\bibitem{gubin2005} A. I. Gubin, K. S. Il'in, S. A. Vitusevich, M. Siegel, and N. Klein, \textit{Dependence of Magnetic Penetration Depth on the Thickness of Superconducting Nb Thin Films}, Phys. Rev. B \textbf{72}, 064503 (2005).
\bibitem{lemberger2007} T. R. Lemberger, I. Hetel, J. W. Knepper, and F. Y. Yang, \textit{Penetration Depth Study of Very Thin Superconducting Nb Films}, Phys. Rev. B \textbf{76}, 094515 (2007).

\bibitem{matsushita} T. Matsushita, \textit{Flux Pinning in Superconductors}, 2nd ed. (Springer, 2014).

\bibitem{kuznetsov1999} A. V. Kuznetsov, D. V. Eremenko, and V. N. Trofimov, \textit{Onset of Flux Penetration into a Thin Superconducting Film Strip}, Phys. Rev. B \textbf{59}, 1507 (1999).
\bibitem{brandt2013} E. H. Brandt, G. P. Mikitik, and E. Zeldov, \textit{Two Regimes of Vortex Penetration into Platelet-Shaped Type-II Superconductors}, J. Exp. Theor. Phys., \textbf{117}, 439 (2013).

\bibitem{bean1964} C. P. Bean and J. D. Livingston, \textit{Surface Barrier in Type-II Superconductors}, Phys. Rev. Lett. \textbf{12}, 14 (1964).

\bibitem{goppl2008} M. G{\" o}ppl, A. Fragner, M. Baur, R. Bianchetti, S. Filipp, J. M. Fink, P. J. Leek, G. Puebla, L. Steffen, and A. Wallraff, \textit{Coplanar Waveguide Resonators for Circuit Quantum Electrodynamics}, J. Appl. Phys. \textbf{104}, 113904 (2008).
\bibitem{sage2011} J. M. Sage, V. Bolkhovsky, W. D. Oliver, B. Turek, and P. B. Welander, \textit{Study of Loss in Superconducting Coplanar Waveguide Resonators}, J. Appl. Phys. \textbf{109}, 063915 (2011).
\bibitem{goetz2016} J. Goetz, F. Deppe, M. Haeberlein, F. Wulschner, C. W. Zollitsch, S. Meier, M. Fischer, P. Eder, E. Xie, K. G. Fedorov, E. P. Menzel, A. Marx, and R. Gross, \textit{Loss Mechanisms in Superconducting Thin Film Microwave Resonators}, J. Appl. Phys. \textbf{119}, 015304 (2016).

\bibitem{coffey1991} M. W. Coffey and J. R. Clem, \textit{Unified Theory of Effects of Vortex Pinning and Flux Creep upon the rf Surface Impedance of Type-II Superconductors}, Phys. Rev. Lett. \textbf{67}, 386 (1991).
\bibitem{brandt1991} E. H. Brandt, \textit{Penetration of Magnetic ac Fields into Type-II Superconductors}, Phys. Rev. Lett. \textbf{67}, 2219 (1991).
\bibitem{dulcic1993b} A. Dul\v ci\'c and M. Po\v zek, \textit{Microwave Surface Impedance in the Mixed State of Type-II Superconductors}, Physica C \textbf{218}, 449 (1993).
\bibitem{pompeo2008} N. Pompeo and E. Silva, \textit{Reliable Determination of Vortex Parameters from Measurements of the Microwave Complex Resistivity}, Phys. Rev. B \textbf{78}, 094503 (2008).
\bibitem{silva2006} E. Silva, N. Pompeo, S. Sarti, and C. Amabile, \textit{Vortex State Microwave Response in Superconducting Cuprates and MgB$_2$}, in \textit{Recent Developments in Superconductivity Research}, edited by B. P. Martins (Nova Science Publishers, New York, 2007).

\bibitem{kopnin} N. B. Kopnin, \textit{Theory of Nonequilibrium Superconductivity} (Oxford University Press, 2001).

\bibitem{song2009a} C. Song, T. W. Heitmann, M. P. DeFeo, K. Yu, R. McDermott, M. Neeley, J. M. Martinis, and B. L. T. Plourde, \textit{Microwave Response of Vortices in Superconducting Thin Films of Re and Al}, Phys. Rev. B \textbf{79}, 174512 (2009).

\bibitem{schuster1994a} Th. Schuster, M. V. Indenbom, H. Kuhn, E. H. Brandt, and M. Konczykowski, \textit{Flux Penetration and Overcritical Currents in Flat Superconductors with Irradiation-Enhanced Edge Pinning: Theory and Experiment}, Phys. Rev. Lett. \textbf{73}, 1424 (1994).
\bibitem{zeldov1994b} E. Zeldov, A. I. Larkin, V. B. Geshkenbein, M. Konczykowski, D. Majer, B. Khaykovich, V. M. Vinokur, and H. Shtrikman, \textit{Geometrical Barriers in High-Temperature Superconductors}, Phys. Rev. Lett. \textbf{73}, 1428 (1994).
\bibitem{zeldov1994c} E. Zeldov, A. I. Larkin, M. Konczykowski, B. Khaykovich, D. Majer, V. B. Geshkenbein, and V. M. Vinokur, \textit{Geometrical Barriers in Type-II Superconductors}, Physica C \textbf{235-240}, 2761 (1994).
\bibitem{schuster1994b} Th. Schuster, H. Kuhn, E. H. Brandt, M. Indenbom, M. R. Koblischka, and M. Konczykowski, \textit{Flux Motion in Thin Superconductors with Inhomogeneous Pinning}, Phys. Rev. B \textbf{50}, 16684 (1994).
\bibitem{maksimov1995} I. L. Maksimov and A. A. Elistratov, \textit{Edge Barrier and Structure of the Critical State in Superconducting Thin Films}, JETP Lett. \textbf{61}, 208 (1995).
\bibitem{willa2014} R. Willa, V. B. Geshkenbein, and G. Blatter, \textit{Suppression of Geometric Barrier in Type-II Superconducting Strips}, Phys. Rev. B \textbf{89}, 104514 (2014).

\bibitem{brandt1993b} E. H. Brandt and M. Indenbom, \textit{Type-II-Superconductor Strip with Current in a Perpendicular Magnetic Field}, Phys. Rev. B \textbf{48}, 12893 (1993).
\bibitem{zeldov1994a} E. Zeldov, J. R. Clem, M. McElkesh, and M. Darwin, \textit{Magnetization and Transport Currents in Thin Superconducting Films}, Phys. Rev. B \textbf{49}, 9802 (1994).

\bibitem{koch1974} C. C. Koch, J. O. Scarbrough, and D. M. Kroeger, \textit{Effects of Interstitial Oxygen on the Superconductivity of Niobium}, Phys. Rev. B \textbf{9}, 888 (1974).
\bibitem{halbritter2005} J. Halbritter, \textit{Transport in Superconducting Niobium Films for Radio Frequency Applications}, J. Appl. Phys. \textbf{97}, 083904 (2005).


\bibitem{mattis1958} D. C. Mattis and J. Bardeen, \textit{Theory of the Anomalous Skin Effect in Normal and Superconducting Metals}, Phys. Rev. \textbf{111}, 412 (1958).
\bibitem{zimmermann1991} W. Zimmermann, E. H. Brandt, M. Bauer, E. Seider, and L. Genzel, \textit{Optical Conductivity of BCS Superconductors with Arbitrary Purity}, Physica C \textbf{183}, 99 (1991).
\bibitem{dressel} M. Dressel, \textit{Electrodynamics of Metallic Superconductors}, Adv. Condens. Matter Phys. \textbf{2013}, 104379 (2013).
\bibitem{white1964} R. H. White and M. Tinkham, \textit{Magnetic-Field Dependence of Microwave Absorption and Energy Gap in Superconducting Films}, Phys. Rev. \textbf{136}, A203 (1964).
\bibitem{maki} K. Maki, \textit{Gapless Superconductivity}, in \textit{Superconductivity}, edited by R. D. Parks (Marcel Dekker, New York, 1969), Vol. 2, p. 1035.
\bibitem{usadel1970} K. D. Usadel, \textit{Generalized Diffusion Equation for Superconducting Alloys}, Phys. Rev. Lett. \textbf{25}, 507 (1970).
\bibitem{anthore2003} A. Anthore, H. Pothier, and D. Esteve, \textit{Density of States in a Superconductor Carrying a Supercurrent}, Phys. Rev. Lett. \textbf{90}, 127001 (2003).


\bibitem{stan2004} G. Stan, S. B. Field, and J. M. Martinis, \textit{Critical Field for Complete Vortex Expulsion from Narrow Superconducting Strips}, Phys. Rev. Lett. \textbf{92}, 097003 (2004).
\bibitem{kuit2008} K. H. Kuit, J. R. Kirtley, W. van der Veur, C. G. Molenaar, F. J. G. Roesthuis, A. G. P. Troeman, J. R. Clem, H. Hilgenkamp, H. Rogalla, and J. Flokstra, \textit{Vortex Trapping and Expulsion in Thin-Film YBa$_2$Cu$_3$O$_{7-\delta}$ Strips}, Phys. Rev. B \textbf{77}, 134504 (2008).

\bibitem{zerweck1981} G. Zerweck, \textit{On Pinning of Superconducting Flux Lines by Grain Boundaries}, J. Low Temp. Phys. \textbf{42}, 1 (1981).

\bibitem{geim2000} A. K. Geim, S. V. Dubonos, I. V. Grigorieva, K. S. Novoselov, F. M. Peeters, and V. A. Schweigert, \textit{Non-Quantized Penetration of Magnetic Field in the Vortex State of Superconductors}, Nature \textbf{407}, 55 (2000).
\bibitem{peeters2002} F. M. Peeters, V. A. Schweigert, and B.J. Baelus, \textit{Fractional and Negative Flux Penetration in Mesoscopic Superconducting Disks}, Physica C \textbf{369}, 158 (2002).

\bibitem{fetter1980} A. L. Fetter, \textit{Flux Penetration in a Thin Superconducting Disk}, Phys. Rev. B \textbf{22}, 1200 (1980).
\bibitem{kogan1994} V. G. Kogan, \textit{Pearl's Vortex Near the Film Edge}, Phys. Rev. B \textbf{49}, 15874 (1994).
\bibitem{pearl1964} J. Pearl, \textit{Current Distribution in Superconducting Films Carrying Quantized Fluxoids}, Appl. Phys. Lett. \textbf{5}, 65 (1964).


\bibitem{sharvin1961} Yu. V. Sharvin and V. F. Gantmakher, \textit{Dependence of the Depth of Penetration of the Magnetic Field in a Superconductor on the Magnetic Field Strength}, Sov. Phys. JETP \textbf{12}, 866 (1961).
\bibitem{douglass1961} D. H. Douglass, Jr., \textit{Magnetic Field Dependence of the Superconducting Penetration Depth in Thin Specimens}, Phys. Rev. \textbf{124}, 735 (1961).
\bibitem{sridhar1989} S. Sridhar, D.-H. Wu, and W. Kennedy, \textit{Temperature Dependence of Electrodynamic Properties of YBa$_2$Cu$_3$O$_y$ Crystals}, Phys. Rev. Lett. \textbf{63}, 1873 (1989).
\bibitem{healey2008} J. E. Healey, T. Lindstr{\"o}m, M. S. Colclough, C. M. Muirhead, and A. Ya. Tzalenchuk, \textit{Magnetic Field Tuning of Coplanar Waveguide Resonators}, Appl. Phys. Lett. \textbf{93}, 043513 (2008).

\bibitem{clem2011} J. R. Clem and K. K. Berggren, \textit{Geometry-Dependent Critical Currents in Superconducting Nanocircuits}, Phys. Rev. B \textbf{84}, 174510 (2011).


\bibitem{klein1990} N. Klein, H. Chaloupka, G. M\"uller, S. Orbach, H. Piel, B. Roas, L. Schultz, U. Klein, and M. Peiniger, \textit{The Effective Microwave Surface Impedance of High $T_c$ Thin Films}, J. Appl. Phys. \textbf{67}, 6940 (1990).
\bibitem{irz1995} D. Yu. Irz, V. N. Ryzhov, and E. E. Tareyeva, \textit{Vortex-Vortex Interaction in a Superconducting Film of Finite Thickness}, Phys. Lett. A \textbf{207}, 374 (1995).
\bibitem{wei1996} J.-C. Wei and T.-J. Yang, \textit{Current Distribution and Vortex-Vortex Interaction in a Superconducting Film of Finite Thickness}, Jpn. J. Appl. Phys. \textbf{35}, 5696 (1996).





\bibitem{schmid1966} A. Schmid, \textit{A Time Dependent Ginzburg--Landau Equation and its Application to the Problem of Resistivity in the Mixed State}, Phys. kondens. Materie \textbf{5}, 302 (1966).
\bibitem{gropp1996} W. D. Gropp, H. G. Kaper, G. K. Leaf, D. M. Levine, M. Palumbo, and V. M. Vinokur, \textit{Numerical Simulation of Vortex Dynamics in Type-II Superconductors}, J. Comput. Phys. \textbf{123}, 254 (1996).
\bibitem{alstrom2011} T. S. Alstr{\o}m, M. P. S{\o}rensen, N. F. Pedersen, and S. Madsen, \textit{Magnetic Flux Lines in Complex Geometry Type-II Superconductors Studied by the Time Dependent Ginzburg-Landau Equation}, Acta. Appl. Math. \textbf{115}, 63 (2011).

\bibitem{sheen1991} D. M. Sheen, S. M. Ali, D. E. Oates, R. S. Withers, and J. A. Kong, \textit{Current Distribution, Resistance, and Inductance for Superconducting Strip Transmission Lines}, IEEE Trans. Appl. Supercond. \textbf{1}, 108 (1991).
\bibitem{chang1979} W. H. Chang, \textit{The Inductance of a Superconducting Strip Transmission Line}, J. Appl. Phys. \textbf{50}, 8129 (1979).

\bibitem{SC1} W. Kern, \textit{The Evolution of Silicon Wafer Cleaning Technology}, J. Electrochem. Soc. \textbf{137}, 1887 (1990).
\bibitem{wenner2011} J. Wenner, R. Barends, R. C. Bialczak, Yu Chen, J. Kelly, E. Lucero, M. Mariantoni, A. Megrant, P. J. J. O’Malley, D. Sank, A. Vainsencher, H. Wang, T. C. White, Y. Yin, J. Zhao, A. N. Cleland, and J. M. Martinis, \textit{Surface Loss Simulations of Superconducting Coplanar Waveguide Resonators}, Appl. Phys. Lett. \textbf{99}, 113513 (2011).
\bibitem{megrant2012} A. Megrant, C. Neill, R. Barends, B. Chiaro, Yu Chen, L. Feigl, J. Kelly, E. Lucero, M. Mariantoni, P. J. J. O’Malley, D. Sank, A. Vainsencher, J. Wenner, T. C. White, Y. Yin, J. Zhao, C. J. Palmstr{\o}m, J. M. Martinis, and A. N. Cleland, \textit{Planar Superconducting Resonators with Internal Quality Factors above one Million}, Appl. Phys. Lett. \textbf{100}, 113510 (2012).
\bibitem{eom2001} C.-B. Eom and J. M. Murduck, \textit{Synthesis and Characterization of Superconducting Thin Films}, Thin Films \textbf{28}, 227 (2001).
\bibitem{wagner1996} T. Wagner, M. Lorenz, and M. R\"uhle, \textit{Thermal Stability of Nb Thin Films on Sapphire}, J. Mater. Res. \textbf{11}, 1255 (1996).
\bibitem{masek1998} K. Ma\v sek and V. Matol\'in, \textit{RHEED Study of Nb Thin Film Growth on $\alpha$-Al$_2$O$_3$ $(0001)$ Substrate}, Thin Solid Films \textbf{317}, 183 (1998).

\bibitem{wildes} A. R. Wildes, J. Mayer, and K. Theis-Br{\"o}hl, \textit{The Growth and Structure of Epitaxial Niobium on Sapphire}, Thin Solid Films \textbf{401}, 7 (2001).

\bibitem{keithley} Keithley, \textit{Low Level Measurements Handbook}, 7th ed. (2013).


\end{thebibliography}
\end{document}